\documentclass[journal=jacsat,manuscript=article]{achemso}

\usepackage{amsmath}
\usepackage{amsmath,amssymb}
\usepackage{graphicx}
\usepackage{caption}
\usepackage{color}
\usepackage{dcolumn}
\usepackage{bm}
\usepackage{float}
\usepackage{subfig}
\usepackage{soul}
\usepackage{comment}

\usepackage{makecell}
\usepackage{multirow,array} 
\usepackage{arydshln}

\usepackage{rotating}

\newcommand{\rateunit}{cm$^3$ molecule$^{-1}$ s$^{-1}$ }
\author{Juan Carlos San Vicente Veliz}
\affiliation{Department of Chemistry, University of Basel,
Klingelbergstrasse 80, CH-4056 Basel, Switzerland}

\author{Debasish Koner} 
\affiliation{Department of Chemistry, University of Basel,
Klingelbergstrasse 80, CH-4056 Basel, Switzerland}

\author{Max Schwilk} \affiliation{Department of Chemistry, University
  of Basel, Klingelbergstrasse 80, CH-4056 Basel, Switzerland}
\altaffiliation{University of Vienna, Faculty of Physics, 1090 Vienna,
  Austria}

\author{Raymond J. Bemish} \affiliation{Air Force Research Laboratory,
  Space Vehicles Directorate, Kirtland AFB, New Mexico 87117, USA}

\author{Markus Meuwly} \affiliation{Department of Chemistry,
  University of Basel, Klingelbergstrasse 80, CH-4056 Basel,
  Switzerland} \altaffiliation{Department of Chemistry, Brown
  University, RI, USA} \email{m.meuwly@unibas.ch}

\title{The C($^{3}$P) + O$_{2}$($^3\Sigma_{g}^{-}$) $\leftrightarrow$
  CO$_2$ $\leftrightarrow$ CO($^{1}\Sigma^{+}$)+ O($^{1}$D)/O($^{3}$P)
  Reaction: Thermal and Vibrational Relaxation Rates from 15 K to
  20000 K}

\begin{document}

\date{\today}

\begin{abstract}
Thermal rates for the C($^{3}$P) + O$_{2}$($^3\Sigma_{g}^{-}$)
$\leftrightarrow$ CO($^{1}\Sigma^{+}$)+ O($^{1}$D)/O($^{3}$P) reaction
are investigated over a wide temperature range based on quasi
classical trajectory (QCT) simulations on 3-dimensional, reactive
potential energy surfaces (PESs) for the $^1$A$'$, $(2)^1$A$'$,
$^1$A$''$, $^3$A$'$ and $^3$A$''$ states. The forward rate matches
measurements at 15 K to 295 K whereas the equilibrium constant
determined from the forward and reverse rates are consistent with
those derived from statistical mechanics at high
temperature. Vibrational relaxation, O+CO($\nu=1, 2$) $\rightarrow$
O+CO$(\nu=0)$, is found to involve both, non-reactive and reactive
processes. The contact time required for vibrational relaxation to
take place is $\tau \geq 150$ fs for non-reacting and $\tau \geq 330$
fs for reacting (oxygen atom exchange) trajectories and the two
processes are shown to probe different parts of the global potential
energy surface. In agreement with experiments, low collision energy
reactions for the C($^{3}$P) + O$_{2}$($^3\Sigma_{g}^{-}$, $v=0$)
$\rightarrow$ CO($^{1}\Sigma^{+}$)+ O($^{1}$D) lead to
CO($^{1}\Sigma^{+}$, $v'=17$) with an onset at $E_{\rm c} \sim 0.15$
eV, dominated by the $^1$A$'$ surface with contributions from the $^3$A$'$
surface. Finally, the barrier for the CO$_{\rm A}$($^{1}\Sigma^{+}$)
+ O$_{\rm B}$($^{3}$P) $\rightarrow$ CO$_{\rm B}$($^{1}\Sigma^{+}$) +
O$_{\rm A}$($^{3}$P) atom exchange reaction on the $^3$A$'$ PES yields
a barrier of $\sim 7$ kcal/mol (0.300 eV), consistent with an
experimentally reported value of 6.9 kcal/mol (0.299 eV).
\end{abstract}
\maketitle

\section{Introduction}
\begin{figure}[h!]
    \centering \includegraphics[scale=0.99]{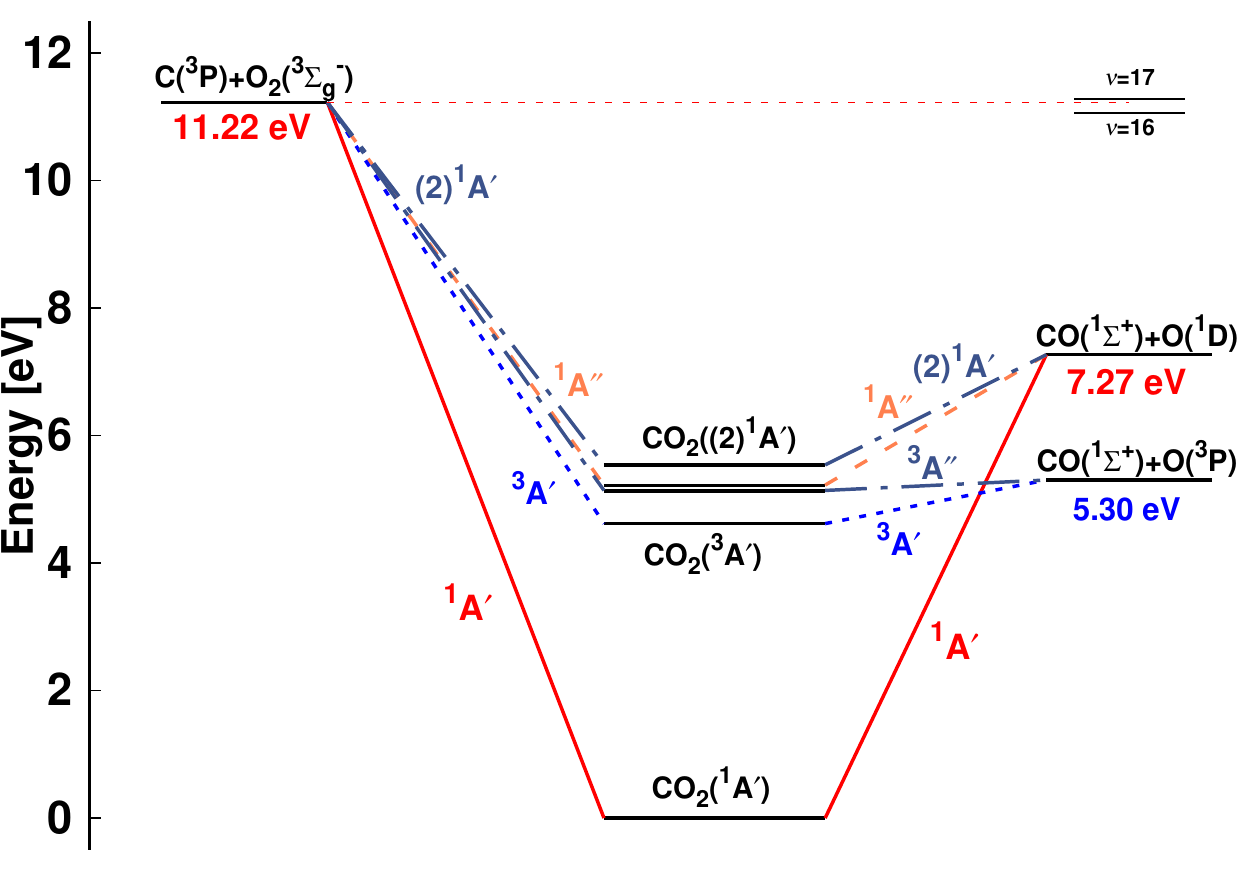}
    \caption{Energy level diagram for the C$+$O$_{2}$
      $\leftrightarrow$ CO$+$O reaction:
      C($^3$P)$+$O$_{2}$($^{3}\Sigma_{g}^{-}$) $\leftrightarrow$
      CO($^1\Sigma^{+}$)$+$O($^3$P) and
      C($^3$P)$+$O$_{2}$($^{3}\Sigma_{g}^{-}$) $\leftrightarrow$
      CO($^1\Sigma^{+}$)$+$O($^1$D). The energies of the dissociating
      species are reported: the O($^1$D)/O($^3$P) separation is 1.97
      eV, consistent with experiment, and the total energies for
      CO$_2$ are $^1$A$'$ ($^1 \Sigma_g$ in D$_{\infty \rm h}$) (0
      eV), $^3$A$'$ (4.62 eV), $^3$A$''$ (5.14 eV), $^1$A$''$ (5.22
      eV), and $(2)^1$A$'$ (5.53 eV). The relative positions of the
      CO$(v'=16)$+O($^1$D) and CO$(v'=17)$+O($^1$D) asymptotes,
      relevant for discussing the low energy collision
      C($^3$P)$+$O$_{2}$($^{3}\Sigma_{g}^{-}$) $\leftrightarrow$
      CO($^1\Sigma^{+}$)$+$O($^1$D) reaction,\cite{costes:1998} are
      indicated on the right hand side. The correlation of the
      $(2)^1$A$'$ state of CO$_2$ based on state-averaged SA-CASSCF
      calculations given in Figure S1 with the
      reactant and product state is consistent with earlier
      work\cite{chastaing:1999} but differs from
      others.\cite{Xantheas1994}}
    \label{fig:fig1}
\end{figure}

Reactions involving carbon and oxygen atoms play important roles in
combustion, hypersonic flow, and planetary
atmospheres.\cite{Sharma2010} Among those, the thermal rates for the
C($^3$P)+O$_2$($^3 \Sigma_{\rm g}^-$), O($^3$P)+CO($^1 \Sigma^+$), and
O($^1$D)+CO($^1 \Sigma^+$) reactions going through various electronic
states of CO$_2$ (see Figure \ref{fig:fig1}) are particularly
relevant. Similarly, the vibrational deactivation of CO($^1 \Sigma^+$)
through collisions with O($^3$P) is a relevant pathway for relaxation
and redistribution of energy in nonequilibrium
flow.\cite{Lewittes1978}\\

\noindent
Several independent studies have determined thermal rates for the
forward C($^3$P) + O$_2$($^3 \Sigma_{\rm g}^-$)
reaction.\cite{husain:1975,bergeat:1999,Geppert2000} Using the CRESU
(Cin\'{e}tique de R\'{e}action en Ecoulement Supersonique Uniforme)
technique\cite{Geppert2000} the thermal rate from experiments between
15 and 295 K was measured. At 298 K the rate was $4.8 \pm 0.5 \cdot
10^{-11}$\rateunit which is within a factor of two to three of other,
previous
experiments.\cite{husain:1975,becker:1988,dorthe:1991,bergeat:1999} In
all three laval nozzle experiments it was found that the rate
increases with decreasing temperature between 15 and 295
K.\cite{Geppert2000,chastaing:1999,chastaing:2000} The product
detection techniques included vacuum ultraviolet laser-induced
fluorescence\cite{Geppert2000,chastaing:2000}, and
chemiluminescence\cite{chastaing:1999}.\\

\noindent
Shock tube experiments of the C+O$_2$ reactions were also carried out
at higher temperatures (from 1500 to 4200 K) and reported a rate of
$k_f(T) = 1.2 \times 10^{14} \exp{(-2010 {\rm K}/T)}$ cm$^{3}$
mol$^{-1}$ s$^{-1}$ (corresponding to $1.9 \times 10^{-10}\exp{(-2010
  {\rm K}/T)}$ \rateunit) with an overall uncertainty of $\pm 50$ \%
and a standard deviation for the activation energy of $\pm 15$ \% and
$\pm 13$ \%, respectively.\cite{Hanson91} Yet earlier emission spectra
in a discharge flow found that the C($^{3}$P) +
O$_{2}$($^3\Sigma_{g}^{-}$) reaction generates CO in high
vibrationally excited states (up to $v' = 17$) and that the transition
state has the configuration COO rather than OCO.\cite{thrush:1973}
Such a COO intermediate was also proposed from the interpretation of
the C+O$_2$ reaction\cite{dubrin:1964} and has been described in
multiconfiguration SCF calculations.\cite{Xantheas1994} Also, no
evidence was found that the C+O$_2$ reaction passes through the region
where the quenching of O($^1$D) to O($^3$P) by CO occurs as a
non-adiabatic process, as had been proposed
earlier.\cite{husain:1973.1,husain:1973.2}\\

\noindent
For the reverse reactions, O($^3$P)+CO($^1 \Sigma^+$), and
O($^1$D)+CO($^1 \Sigma^+$) leading to C($^{3}$P) +
O$_{2}$($^3\Sigma_{g}^{-}$), the onset for the rates $k_r(T)$ to form
C+O$_2$ is expected to occur at considerably higher temperature than
that for $k_f$ due to the large energy difference of $\sim 6$ eV
between the O+CO and the C+O$_2$ asymptotes, see Figure
\ref{fig:fig1}. There are, however, computational investigations of
the oxidation of CO to form CO$_2$ following the O($^3$P) + CO($^1
\Sigma^+$) $\rightarrow$ CO$_2$($^1 \Sigma_{\rm g}^+$) route, usually
involving a third particle M.\cite{Jasper2013} The rates for formation
of CO$_2$ along the $^3$A$'$ and $^3$A$''$ pathways starting from
O($^3$P)+CO($^1 \Sigma^+$) ranged from $10^{-13}$ to $10^{-12}$ cm$^3$
molecule$^{-1}$ s$^{-1}$, depending on temperature, compared with
$\sim 10^{-14}$ cm$^3$ molecule$^{-1}$ s$^{-1}$ from earlier
work.\cite{troe:1975} These were non-Born-Oppenheimer dynamics
simulations of the O($^3$P)+CO($^1 \Sigma^+$) $\rightarrow$ CO$_2$($^1
\Sigma_{\rm g}^+$) reaction involving the $^1$A$'$, $^3$A$'$, and
$^3$A$''$ potential energy surfaces (PESs).\cite{Jasper2013} The
spin-forbidden fraction in this study was, however, found to be small
($\sim 1$ \%). Experimentally, the forward reaction has not been
probed so far, to the best of our knowledge. Direct experiments
involving [O($^3$P), O($^1$D)] and CO($^1 \Sigma^+$) concern the
vibrational deactivation of CO upon collision with atomic
oxygen.\cite{Braunstein2000,shatalov:2000,Lewittes1978,Center1973,Kelley1977}
Finally, the rate for collisional spin relaxation for the O($^1$D) to
O($^3$P) spin relaxation by CO($^1 \Sigma^+$) at temperatures between
113 and 333 K was determined.\cite{Davidson1978} The rates were found
to vary monotonically from about $7.6 \times 10^{-11}$ to $5.2 \times
10^{-11}$ cm$^3$ molecule$^{-1}$ s$^{-1}$ over the temperature
range. Earlier modeling based on collisions with CO and other small
molecules obtained a rate of $8 \times 10-11$ cm$^3$ molecule$^{-1}$
s$^{-1}$.\cite{Tully:1975}\\

\noindent
Computationally, the ground and excited state PESs for CO$_2$ have
been studied in some
detail.\cite{winter:1973,Xantheas1994,greben:2012,greben:2013,greben:2013.2,Schmidt2013,hsien:2013}
Early configuration interaction calculations
established\cite{winter:1973} that there must be four states (two
singlet and two triplet) of CO$_2$ below the
CO($^1\Sigma^{+}$)$+$O($^3$P) asymptote which is also what is found in
the present work (Figure \ref{fig:fig1}). CO$_2$ does not show strong
absorptions below 11 eV\cite{winter:1973} which makes direct
comparison difficult also, because often vertical and not adiabatic
transition energies were measured. A low-lying adiabatic electronic
transition to a triplet state was reported at 39412 cm$^{-1}$ (4.89
eV) above the ground state,\cite{Dixon1963} in qualitative agreement
with the position of the $^3$A$'$ state, 4.62 eV above the ground
state, see Figure \ref{fig:fig1}.\\

\noindent
An early classical MD study\cite{murrell:1977} of the forward reaction
using an analytical potential energy surface found a rate of $k_f =
1.92 \times 10^{-11}$ cm$^3$ molecule$^{-1}$ s$^{-1}$.  In dynamics
studies\cite{Braunstein2000,Brunsvold2008,Jasper2013,Schmidt2013,schwenke2016}
the reference energies from electronic structure calculations were
either represented as parametrized
fits,\cite{Braunstein2000,Brunsvold2008,schwenke2016} cubic
splines,\cite{Schmidt2013} or interpolated moving least
squares.\cite{Jasper2013} Reference calculations were carried out at
the CASSCF-MP2/631G+(d),\cite{Braunstein2000} and MRCI+Q/aug-cc-pVQZ
levels of theory.\cite{Jasper2013,Schmidt2013,schwenke2016} The
dynamics simulations either concerned the O-induced collisional
dissociation of CO,\cite{schwenke2016} CO vibrational
relaxation,\cite{Braunstein2000} the O-exchange dynamics in reactive
O+CO collisions, non-Born-Oppenheimer effects in CO$_2$ formation from
O+CO collisions,\cite{Jasper2013} or the final state distributions
from the O+CO reactive scattering\cite{Brunsvold2008} but not the
entire C($^{3}$P) + O$_{2}$($^3\Sigma_{g}^{-}$) $\leftrightarrow$
CO$_2$ $\leftrightarrow$ CO($^{1}\Sigma^{+}$)+ O($^{1}$D)/O($^{3}$P)
reaction involving several electronic states.\\

\noindent
A schematic of the states derived from the present calculations and
considered in the present work is provided in Figure
\ref{fig:fig1}. The left hand side is the C($^{3}$P) +
O$_{2}$($^{3}\Sigma_{g}^{-}$) (entrance) channel which connects to all
bound CO$_2$ states in the middle. This asymptote is 11.22 eV above
the global minimum which is the linear CO$_2$($^1$A$'$) structure. The
right hand side of Figure \ref{fig:fig1} shows the two product
channels considered: the lower CO($^{1}\Sigma^{+}$) + O($^{3}$P)
state, 5.30 eV above the minimum energy of the CO$_2$($^1$A$'$) ground
state, and the CO($^{1}\Sigma^{+}$) + O($^{1}$D) asymptote another
1.97 eV higher in energy. The final state involving O($^{3}$P)
connects with the triplet states ($^3$A$'$ and $^3$A$''$) of CO$_2$
whereas that leading to O($^{1}$D) correlates with the $^1$A$'$,
$^1$A$''$, and $(2) ^1$A$'$ states, see Figure \ref{fig:fig1}.\\

\noindent
Except for the shock tube experiments\cite{Hanson:1991} on C+O$_2$
$\rightarrow$ O+CO (1500-4200 K) and the computations\cite{Jasper2013}
for CO$_2$ formation from O+CO (between 1000 K and 5000 K) there is
little information on the high-temperature dynamics of either, the
C+O$_2$ or the O+CO reactive processes. The present work extends this
by performing QCT simulations on the 5 lowest states of CO$_2$,
represented as a reproducing kernel Hilbert space
(RKHS),\cite{ho96:2584,MM.rkhs:2017} and focusing on the forward and
reverse reactions and vibrational relaxation. First, the methods are
presented and the potential energy surfaces for all 5 states are
discussed. Then the thermal rates are determined along the singlet
pathway. Next, vibrational relaxation for the O+CO collision is
considered for CO$(v=1)$ and CO$(v=2)$ and the distributions for
relaxing/nonrelaxing reactive/nonreactive trajectories are mapped onto
the PES. Finally, conclusions are drawn.\\

\section{Computational Methods}
This section presents the generation and representation of the
potential energy surfaces and the methodologies for the QCT
simulations and their analysis.

\subsection{Electronic Structure Calculations}
All PESs are computed at the multi reference CI singles and doubles
(MRCISD) level of theory\cite{wer88:5803,kno88:514} including the
Davidson quadruples correction\cite{davidson:1974} (MRCISD+Q) together
with the aug-cc-pVTZ basis set\cite{dun89:1007} using the MOLPRO
2019.1 software.\cite{MOLPRO_brief} In order to consistently describe
all relevant states and avoid numerical instabilities due to
closely-lying states of the same symmetry, state-averaged
CASSCF\cite{wen85:5053,kno85:259,wer80:2342,werner:2019} calculations
including the two lowest states of each symmetry (two spin symmetries
and two spatial symmetries) were carried out. Hence, in total eight
states are included in the CASSCF reference wave function. MRCI+Q
calculations for both asymptotic channels followed for the 5 lowest
CO$_2$ states, namely $^1$A$'$, $^3$A$'$, $^3$A$''$, $^1$A$''$, and
$(2)^1$A$'$, see Figure \ref{fig:fig1}.\\

\noindent
The energies were computed on a grid defined by Jacobi coordinates
$(r,R,\theta)$ where $r$ is the separation of the diatomic, $R$ is the
distance between the atom and the center of mass of the diatomic and
$\theta$ is the angle between the two unit vectors $\vec{r}$ and
$\vec{R}$. For channel I (C($^3$P) + O$_2$($^3 \Sigma_g^-$)) the
$R-$grid included 28 points between 1.4 and 11 a$_{0}$ and the
distance $r$ was covered by 20 points between 1.55 and 4.10 a$_{0}$
whereas for channel II (O($^3$P/$^1$D) + CO($^1 \Sigma^+$)) the
$R-$grid included 26 points between 1.8 and 11 a$_{0}$, and the
distance $r$ was covered by 20 points between 1.55 and 4.00
a$_{0}$. The angular grid for both channels contained 13 angles from a
Gauss-Legendre quadrature (169.796$^\circ$, 156.577$^\circ$,
143.281$^\circ$, 129.967$^\circ$, 116.647$^\circ$, 103.324$^\circ$,
90.100$^\circ$, 76.676$^\circ$, 63.353$^\circ$, 50.033$^\circ$,
36.719$^\circ$,23.423$^\circ$, 10.204$^\circ$).\\

\noindent
The reference points are then represented using reproducing kernel
Hilbert space (RKHS)
techniques.\cite{ho96:2584,ho00:3960,MM.rkhs:2017} The quality of the
representation is further checked using energies from additional,
off-grid geometries. The global, reactive 3D PES $V(r_1,r_2,r_3)$ for
an electronic state is constructed by summing the weighted individual
PESs for each channel
\begin{equation}
V(r_1,r_2,r_3)=\sum_{j=1}^{3} w_{j} (r_{j}) V_{j} (R,r_j,\theta),
\label{eq:mixed}
\end{equation}
using an exponential switching function with weights
\begin{equation}
w_{i} (r)=\frac{e^{-(r_{i}/\sigma_{i})^{2}}}{\sum_{j=1}^{3}
  e^{-(r_{j}/\sigma_{j})^{2}}}.
\label{eq:mixed2}
\end{equation}
Here, $\sigma_{i}$ are switching function parameters for channels I
and II. These parameters were optimized by a least squares fit and
yielded values of (0.90, 1.00, 1.00) a$_0$, (1.10, 1.05, 1.05) a$_0$,
(0.9, 1.00, 1.00) a$_0$, (0.85, 1.25, 1.25) a$_0$ and (1.05, 1.00,
1.00) a$_0$ for the $^1$A$'$, $(2)^1$A$'$, $^1$A$''$, $^3$A$'$, and
$^3$A$''$ PESs, respectively.\\

\noindent
The - global and local - minima and transition states between the
minima and/or entrance channels supported by the PESs were determined
using BFGS minimization and the nudged elastic band
method\cite{jonsson:2000} as implemented in the atomic simulation
environment (ASE)\cite{Larsen:2017}.\\

\subsection{Quasi-Classical Trajectory Simulations}
The QCT simulations used in the present work have been extensively
described in the literature\cite{tru79,hen11,kon16:4731,MM.cno:2018}.
Here, Hamilton's equations of motion are solved using a fourth-order
Runge-Kutta method. The time step was $\Delta t = 0.05$ fs which
guarantees conservation of the total energy and angular
momentum. Initial conditions for the trajectories are sampled using
standard Monte Carlo methods.\cite{tru79} The reactant and product
ro-vibrational states are determined following semiclassical
quantization with quantum bound state calculations for the two
diatomics. Since the ro-vibrational states of the product diatom are
continuous numbers, the states need to be assigned to integer values
for which a Gaussian binning (GB) scheme was used. For this, Gaussian
weights centered around the integer values with a full width at half
maximum of 0.1 were used.\cite{bon97:183,bon04:106,kon16:4731} It is
noted that using histogram binning (HB) was found to give comparable
results for a similar system.\cite{MM.cno:2018}\\

\noindent
The thermal rate for an electronic state ($i$) at a given temperature
($T$) is then obtained from
\begin{equation}
 k_i(T) = g_i(T)\sqrt{\frac{8k_{\rm B}T}{\pi\mu}} \pi b^2_{\rm max}
 \frac{N_{r}}{N_{\rm tot}},
\label{eq:thermal}
\end{equation}
where $g_i(T)$ is the electronic degeneracy factor of state `$i$',
$\mu$ is the reduced mass of the collision system, $k_{\rm B}$ is the
Boltzmann constant, and, depending on the specific process considered,
$N_r$ is the number of reactive or vibrationally relaxed
trajectories. In the rate coefficient calculations, the initial
ro-vibrational states and relative translational energy (collision
energy $E_{\rm c}$) of the reactants for the trajectories are sampled
from Boltzmann and Maxwell-Boltzmann distributions at given $T$,
respectively. The sampling methodology is discussed in detail in
Ref. \cite{MM.cno:2018}\\

\noindent
For the forward C($^{3}$P) + O$_{2}$($^3\Sigma_{g}^{-}$) $\rightarrow$
CO($^{1}\Sigma^{+}$)+ O($^{1}$D) and reverse reactions
CO($^{1}\Sigma^{+}$)+ O($^{1}$D) $\rightarrow$ C($^{3}$P) +
O$_{2}$($^3\Sigma_{g}^{-}$) with rates $k_{\rm f}(T)$ and $k_{\rm
  r}(T)$, respectively, the degeneracy factor is
\begin{equation}
g_{(^1A', (2)^1A', ^1A'')}(T)=\frac{1}{(1+3 \cdot
  e^{\frac{-23.6}{T}}+5\cdot e^{\frac{-62.4}{T}})}
\label{eq:ge1}
\end{equation}
where the terms are the degeneracies of the $J$ states for which the
energy differences between the ground and the excited states are
--23.6 K and --62.4 K, respectively. For the reactions leading to
O($^3$P) and going through triplet CO$_2$ the degeneracies are
$g_{(^3A', ^3A'')}=\frac{1}{3}$. From $k_{\rm f}(T)$ and $k_{\rm
  r}(T)$ the equilibrium constant
\begin{equation}
K_{\rm eq}(T)=\frac{k_{\rm f}(T)}{k_{\rm r}(T)}.
\label{eq:equilibrium}
\end{equation}
is determined.

\section{Results and Discussion}
\subsection{The Potential Energy Surfaces}
\begin{figure}
\centering
\includegraphics[scale=0.65]{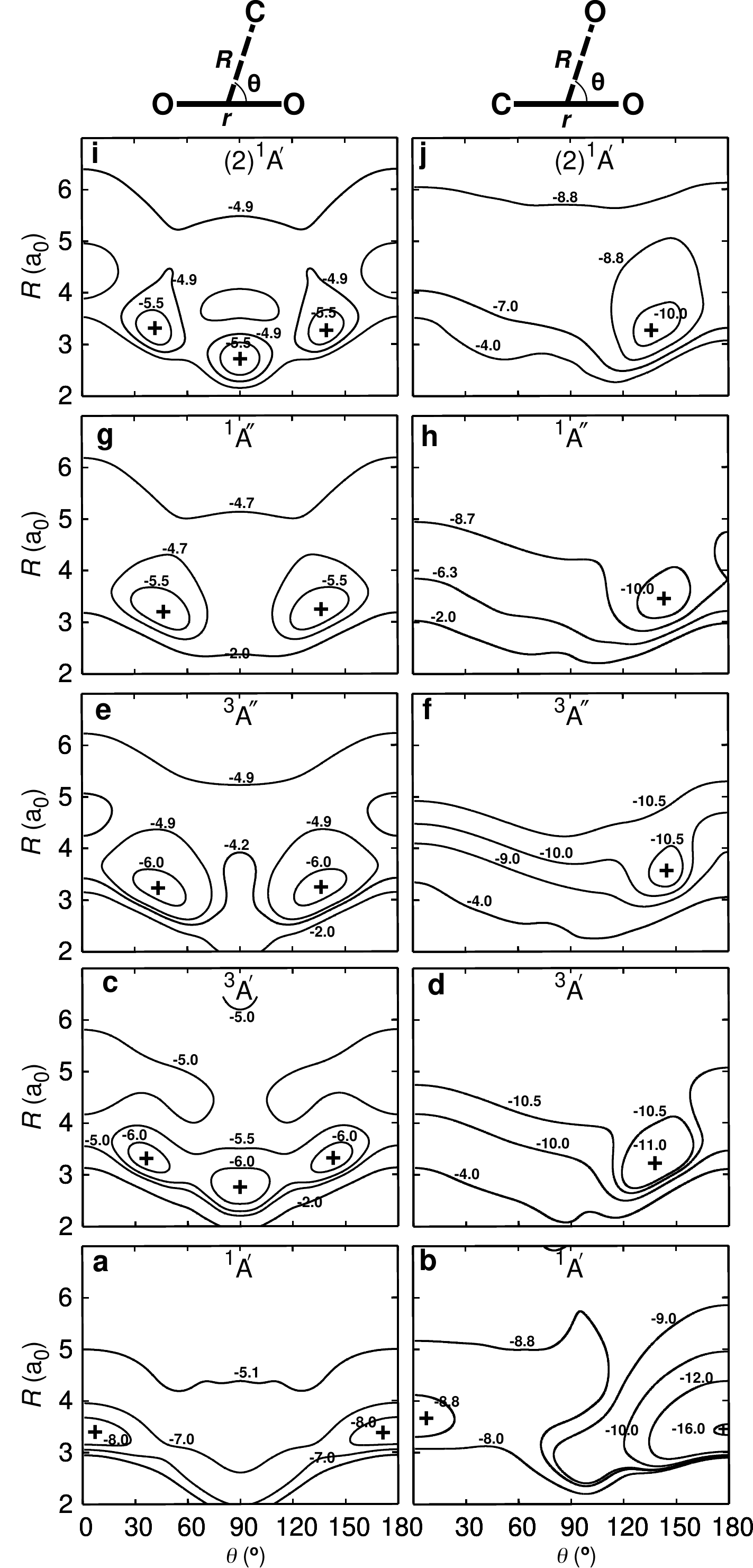}
\caption{Two-dimensional cuts through the 3-d PES for the OO+C (left)
  and the CO+O (right) channels. Energy contours (in eV) for the
  $^1$A$'$ (panels a,b), $^3$A$'$ (panels c,d), $^3$A$''$ (panels
  e,f), $^1$A$''$ (panels g,h), and $(2)^1$A$'$ (panels i,j)
  states. The OO and CO diatomic distance are fixed at $r=2.14$ and
  $r=2.29$ a$_{0}$ for the CO+O and OO+C channels, respectively. The
  zero of energy is the dissociation into atomic fragments (C($^3$P) +
  O($^3$P) + O($^3$P)). The symbol (``+'') indicates local and global
  minima on each PES.}
\label{fig:PES}
\end{figure}

Two-dimensional contour plots of the PESs are shown in Figure
\ref{fig:PES} and the positions and relative energies of the critical
points are summarized in Table 1. The left hand column in Figure
\ref{fig:PES} reports the PESs for the C+O$_2$ asymptote whereas the
right hand column that for the CO+O channel, including the linear
ground state structure for CO$_2$ (Figure \ref{fig:PES}b). All PESs
for the C+O$_2$ asymptote are manifestly symmetric with respect to
$\theta = 90^\circ$ with moderate anisotropies for the $^1$A$'$ and
$^3$A$'$ states and increased anisotropies for all other
PESs. Conversely, all PESs for the O+CO channel are single-minima PESs
with their minima around $140^\circ$ , except for the $^1$A$'$ state
which has a minima for the OCO $(180^\circ)$ and OOC $(0^\circ)$
structures, see also Figure S2. The energy of the OOC
state is 170.0 kcal/mol (7.37 eV) above the OCO minimum and the
barrier height for transition between the OOC and OCO minima is 8.5
kcal/mol (0.369 eV). In addition, the existence of the local OOC
minimum was confirmed at the MRCI+Q and CCSD(T) levels of theory and
was suggested earlier from experiments\cite{dubrin:1964,thrush:1973}
and calculations.\cite{Xantheas1994}\\

\noindent
The quality of the RKHS representations is reported in Figure 
S3. All root mean squared errors for both, on-grid and
off-grid points are below 1 kcal/mol (0.043 eV), except for the
$^1$A$''$ PES for which it is 1.04 kcal/mol (0.045 eV), all evaluated
on the mixed PESs, see Eqs. \ref{eq:mixed} and \ref{eq:mixed2}. For
the individual channels the performance of the RKHS is even
better. One dimensional cuts for an OCO angle of $120^\circ$ directly
comparing the reference points and the RKHS are shown in Figure
S4 for the lowest five states. Importantly, for off-grid
points which were not used to construct the PESs but to independently
validate the RKHS representations, the RMSEs are all below 1 kcal/mol
(0.043 eV), too.\\

\begin{table}[ht]
    \centering
    \resizebox{\textwidth}{!}{
      \caption{Minima (MIN$i$) and transition states (TS$i$) on the
        CO$_2$ PESs towards the CO($^{1}\Sigma^{+}$) + O($^3$P)
        asymptote using the Nudged Elastic Band (NEB)
        method.\cite{jonsson:2000} The PESs are in ascending energetic
        order, see Figure \ref{fig:PES}. Distances are in a$_{0}$ and
        Energy in eV. Energies relative to the CO($^{1}\Sigma^{+}$) +
        O($^3$P) dissociation limit. Present values are compared with
        previous work: for triplet states top row
        Ref.\cite{Jasper2013}, bottom row
        Ref.\cite{Braunstein2000}. For the global minimum in the
        $^1$A$'$ state: top row experiment\cite{NIST}, bottom row
        Ref.\cite{Jasper2013} and for the remaining minima and
        transition states Ref.\cite{Hwang2000}. Where necessary,
        literature values were converted to a$_0$ and eV.}
    \begin{tabular}[c]{m{3 cm}|cccc||m{1.4cm}m{1.4cm}m{1.4cm}m{1.4cm}}
      \hline
           &                                         & Present Work &  &      &  & Literature\cite{Jasper2013,Braunstein2000,Braunstein2009,Hwang2000,NIST} & &  \\
        \hline
    PES & $r_{1}$(CO$_{\rm A}$) & $r_{2}$(CO$_{\rm B}$) & $\angle$(OCO) & $E$ (eV) & $r_{1}$ & $r_{2}$ & $\angle$(OCO) & $E$ (eV) \\
    
    \hline
    $(2)^{1}$A$'$ MIN & 2.368 & 2.368 &  119.0  & 0.23 & -- & -- & -- & -- \\
    \hline
    \hline
    $^{1}$A$''$ MIN & 2.374  &  2.374 &  130.5   & --0.13 & -- & -- & -- & --\\
    \hline
    \hline
     $^{3}$A$''$ MIN & 2.374 & 2.374 &  130.6 & --0.14 & 2.364 2.399  & 2.364 2.399 &  127.2 127.0   & \makecell{--0.23\\--0.22} \\ \cline{6-9} 
     
     $^{3}$A$''$ TS & 2.165 & 3.431 &  135.8 & 0.47 & 2.147 2.192  & 3.515 3.496 & 126.2 122.0   & \makecell{  0.35\\0.30} \\
        \hline
        \hline
        $^{3}$A$'$ MIN &  2.356   &  2.356 &  126.3  & --0.69 & 2.381 2.349   &  2.381 2.349  &  118.0 118.0   &  \makecell{--0.94\\--0.92}   \\ \cline{6-9}
    
    $^{3}$A$'$ TS & 2.163 & 3.544  & 116.3 & 0.39 &  2.143 2.192   &  3.628 3.779  &  120.8 112.0    & \makecell{0.28\\0.20}  \\
    \hline
    \hline
    $^{1}$A$'$ (Global-M) & 2.206 & 2.206 &  180.0 & --5.30 & 2.196 2.194 & 2.196 2.194 & 180.0 180.0  & \makecell{--5.45\\--5.64} \\  \cline{6-9}
        $^{1}$A$'$ (MIN1) & 2.527 & 2.527 &  70.6 & 0.74 & 2.522  & 2.522   & 72.9  & \makecell{0.61} \\
        $^{1}$A$'$ (MIN2) & 2.192 & 4.798 &  0.0 & 1.88 & 2.220 & 4.716 & 0.0  & \makecell{1.72} \\
        $^{1}$A$'$ (TS1) & 2.502 & 2.430 &  88.4 & 1.05 & 2.600 & 2.309 & 91.8  & \makecell{1.04} \\
        $^{1}$A$'$ (TS2) & 2.164 & 4.279 &  68.9 &  2.15  & 2.203 & 4.537 & 41.0  & \makecell{2.22} \\
    \hline
    \hline
    \end{tabular}
   \label{tab:minima}   }
\end{table}

\begin{figure}
\centering \includegraphics[scale=0.04]{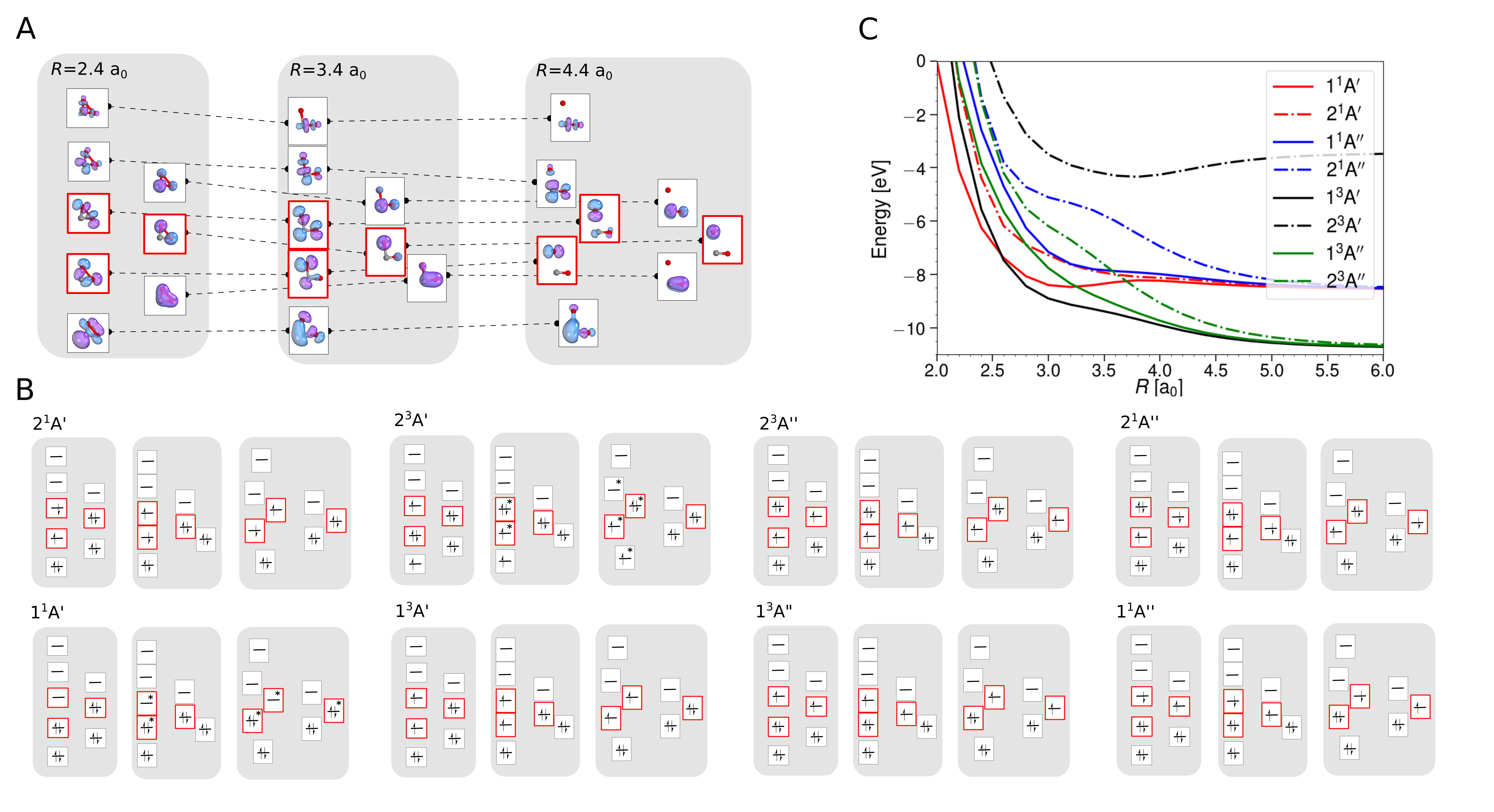}
\caption{Analysis of the all-valence active space
  SA-CASSCF/aug-cc-pVTZ wave functions for a bent geometry with
  $\theta = 117.65^\circ$, $r_{\rm CO} = 2.14$ a$_0$ and for $R=2.4,
  3.4, 4.4$ a$_0$ (CO$_2$ $\rightarrow$ CO + O dissociation path). A:
  Valence molecular orbitals (natural orbitals) energetically close to
  the three frontier orbitals (red frames) whose occupation defines
  the five lowest lying electronic states considered for the dynamics
  on the CO$_2$ PES. The three depicted geometries are oriented with
  the symmetry plane parallel to the paper plane. Orbitals symmetric
  and antisymmetric with respect to the plane in the left and right
  columns of each diagram, respectively. B: Dominant configuration
  state functions of the eight states included in the SA-CASSCF
  calculation depicted as MO diagrams of the orbitals presented in
  panel A. If other configurations contributed with a weight $>0.05$,
  the orbitals involved in the entanglement are marked by an asterisk
  (i. e. these orbitals have an occupation number that deviates
  significantly from the depicted integer value). C: Energy curves of
  the eight states for the CO$_2$ $\rightarrow$ CO + O dissociation at
  this bent geometry. The ground state C + O + O energy computed at
  the same level of theory is the reference energy.}
\label{fig:MO}
\end{figure}

\noindent
For a better understanding of the shapes of the PESs, the
SA-CASSCF/aug-cc-pVTZ wave functions were analyzed in more detail for
the different states at a bent geometry ($\theta = 117.65^\circ$), see
Figure \ref{fig:MO}, for $r_{\rm CO} = 2.14$ a$_0$ and varying
$R$. Figure \ref{fig:MO}A shows the valence molecular orbitals that
are relevant for the description of the eight states in the SA-CASSCF
calculations along the CO$_2$ $\rightarrow$ CO + O dissociation for
this bent geometry. Figure \ref{fig:MO}B depicts the dominant
configuration state functions along this dissociation path. All states
except for the energetically high lying (2)$^3$A$'$ state resolve with
one dominant CASSCF configuration state function for the eight
computed states of SA-CASSCF and keep their characteristic
configuration along the entire dissociation path. Hence, no avoided
crossing of two states with the same symmetry is observed. Figure
\ref{fig:MO}C shows the relative energetics (taking the C($^3$P) +
O($^3$P) + O($^3$P) ground state computed with the same level of
theory as the reference).\\

\noindent
Upon bending, the doubly degenerate $\pi_3$ non-bonding [doubly
  occupied in $^1$A$'$], as well as the $\pi_3$ antibonding
[unoccupied in $^1$A$'$] orbitals for collinear CO$_2$ undergo a
splitting due to the lifted degeneracy. This results in a Jahn-Teller
splitting of the states $^1$A$''$, $^3$A$''$, $^3$A$'$, and
(2)$^1$A$'$ of CO$_2$ with their energy minimum at a bent geometry,
see right hand column in Figure \ref{fig:PES}. The splitting of the
degenerate HOMO and LUMO $\pi_3$ orbitals upon bending leads to three
frontier orbitals, similar in energy, and with overall occupation of
four electrons in all five energetically low-lying states (red frames
in Figure \ref{fig:MO}A). One of these three frontier orbitals has
$\sigma^*$ character along the O-O bond (see Figure \ref{fig:MO}A) and
is somewhat higher in energy. States that involve double occupation of
this orbital lie higher in energy. Along the same line, states that
involve single occupation of one of the strongly bonding orbitals
below the frontier orbitals also lie energetically higher.\\

\noindent
The $^3$A$'$ and $^3$A$''$ states are lower in energy than the
$^1$A$'$ state for certain bent geometries (see Figures \ref{fig:PES}
and \ref{fig:MO}C), as the triplet states gain from increased Pauli
exchange, as well as reduced Coulomb repulsion due to the single
occupations of orbitals. The corresponding open shell singlet states
[(2)$^1$A$'$ and $^1$A$''$] lie slightly higher in energy than their
triplet counterparts due to reduced Pauli exchange.\\

\noindent
All CO$_2$ singlet states connect to CO($^1 \Sigma^+$)+O($^1$D) upon
dissociation whereas the $^3$A$'$, $^3$A$''$, and (2)$^3$A$''$ states
connect to the CO($^1 \Sigma^+$)+O($^3$P) state. On the other hand,
the (2)$^3$A$'$ connects to the energetically high-lying excited
CO$(^3 \Pi)$+O($^3$P) state. The low-lying triplet CO$_2$ states have
no or rather low barriers towards their dissociation across the entire
PES (see Figure \ref{fig:PES} and Figure \ref{fig:MO}C). Specifically,
the (2)$^3$A$''$ state connects to the ground state of CO + O and
crosses the singlet states upon dissociation. The crossing should
nevertheless only lead to negligible non-adiabatic transition rates,
as they are spin-forbidden. Since this state involves double
occupation of an orbital with $\sigma^*$ character of the O-O bond, it
correlates with high lying excited states in the C + O$_2$ channel and
is energetically well separated from the $^3$A$''$ state whenever
there are short O-O distances. It is therefore sufficient to take its
occupation only into account via the degeneracy of O($^3$P) in the
quasi-classical treatment of the CO + O dissociation channel.\\

\subsection{Forward and Reverse Rates and the Equilibrium Constants}
{\it The forward reaction} C($^{3}$P) + O$_{2}$($^3\Sigma_{g}^{-}$)
(Figure \ref{fig:fig1}) generates ground and excited state oxygen
($^3$P and $^1$D). The pathway to yield $^3$P involves the $^3$A$'$
and $^3$A$''$ CO$_2$ PESs whereas that to form $^1$D goes through the
$^1$A$', (2)^1$A$'$, and $^1$A$''$ states. For each of the reactions
on each PES a minimum of $5 \times 10^5$ trajectories was run at each
temperature.\\

\begin{figure}[h]
    \begin{center}
      \includegraphics[scale=0.45]{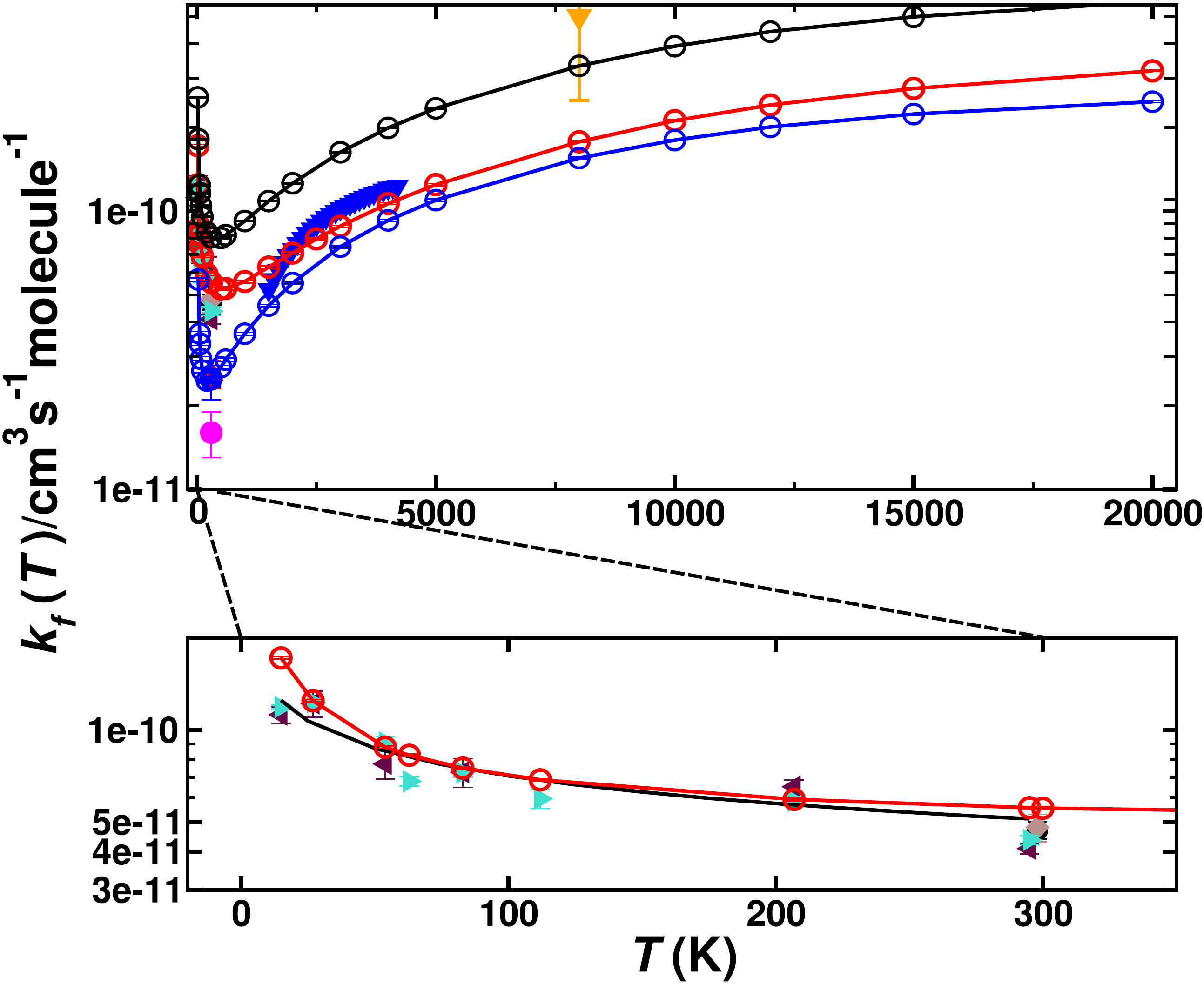}
\caption{Thermal rate for the forward reaction ($k_{\rm f}$) C($^3$P)
  + O$_2$($^3\Sigma_{g}^{-}$) $\rightarrow$ CO($^{1}\Sigma^{+}$) +
  O($^1$D)/O($^{3}$P). The sum of the contribution of the singlet (red
  circles) and triplet (blue circles) states and the total rate (black
  circles). Comparison with forward rates from experiments:
  Ref. \cite{Hanson:1991} (solid blue triangles)
  Ref. \cite{chastaing:1999} (solid green right triangle),
  Ref. \cite{dorthe:1991} (solid magenta circle),
  Ref. \cite{becker:1988} (solid black circle), \cite{BERGEAT1999}
  (solid blue square), Ref. \cite{Geppert2000} (grey diamond),
  Ref. \cite{Fairbairn1969} (solid orange triangle down) and
  Ref. \cite{Husain1975} (red triangle). The bottom panel shows an
  enlarged view for $0 < T < 300$ K for the total singlet rate (solid
  red line) together with the experimental results and a fit using
  Arrhenius parameters provided in the literature\cite{Geppert2000}
  (inset, black solid line).}
\label{fig:cpo2rate}
\end{center}
\end{figure}

\noindent
Figure \ref{fig:cpo2rate} shows the total thermal rates for formation
of O($^1$D) and O($^3$P). The rates for formation of O($^1$D) start at
$1.72 \times 10^{-10}$ \rateunit at 15 K, drop to $5.19
\times10^{-11}$ \rateunit for $T \sim 600$ K and then monotonically
increase to $3.23 \times 10^{-10}$ \rateunit for higher temperatures,
see red line for the total rate in Figure \ref{fig:cpo2rate} with
explicit numerical values reported in Table S1
which also reports the number of reactive trajectories that contribute
to the rate. Experimentally, the total rate for this process was
determined over the temperature range from 1500 K to 4000
K. \cite{Hanson:1991} Evaluating the reported expression $k(T) = 1.2
\times 10^{14}e^{\frac{-2010 K}{T}} (\pm 50 \%)$ \rateunit at 1500 K
and 4000 K yields rates of $k(1500)=5.22 \times 10^{-11}$ \rateunit
and $k(4000)=1.21 \times 10^{-10}$ \rateunit. This compares with
$k(1500)=5.96 \times 10^{-11}$ \rateunit and $k(4000)=1.05 \times
10^{-10}$ \rateunit, respectively, from the present simulations. A
high temperature measurement at 8000 K, associated with a substantial
uncertainty, yields $k \sim 5 \times 10^{-10}$
\rateunit.\cite{Fairbairn1969}\\

\noindent
The inset of Figure \ref{fig:cpo2rate} reports the low-temperature
results. Starting at 15 K the rate first decreases, goes through a
minimum (at $\sim 600$ K) before it raises again for higher
temperatures. Such a behaviour is indicative of a submerged
barrier\cite{sims:2013} which, based on the rates for individual
surfaces, appears to be dominated by the $^1$A$'$ and $^3$A$'$ states,
as seen in Figure S5. Compared with experiments all
computed rates are within 2 \% to 20 \% at 50 K and 30 \% to 40 \% for
Ref. \cite{chastaing:2000} and 4 \% to 30 \% at 300 K for
Ref.\cite{Geppert2000} which can be considered good agreement. For the
process to yield O($^3$P) the individual rates from the contribution
of both triplet PESs ($^3$A$'$ and $^3$A$''$) as well as the total
weighted sum from the process to yield O($^1$D) and O($^3$P) are also
reported in Figure \ref{fig:cpo2rate} (blue and black lines), with
numerical values given in Table S2.\\

\begin{figure}[h]
    \begin{center}
    \includegraphics[scale=0.55]{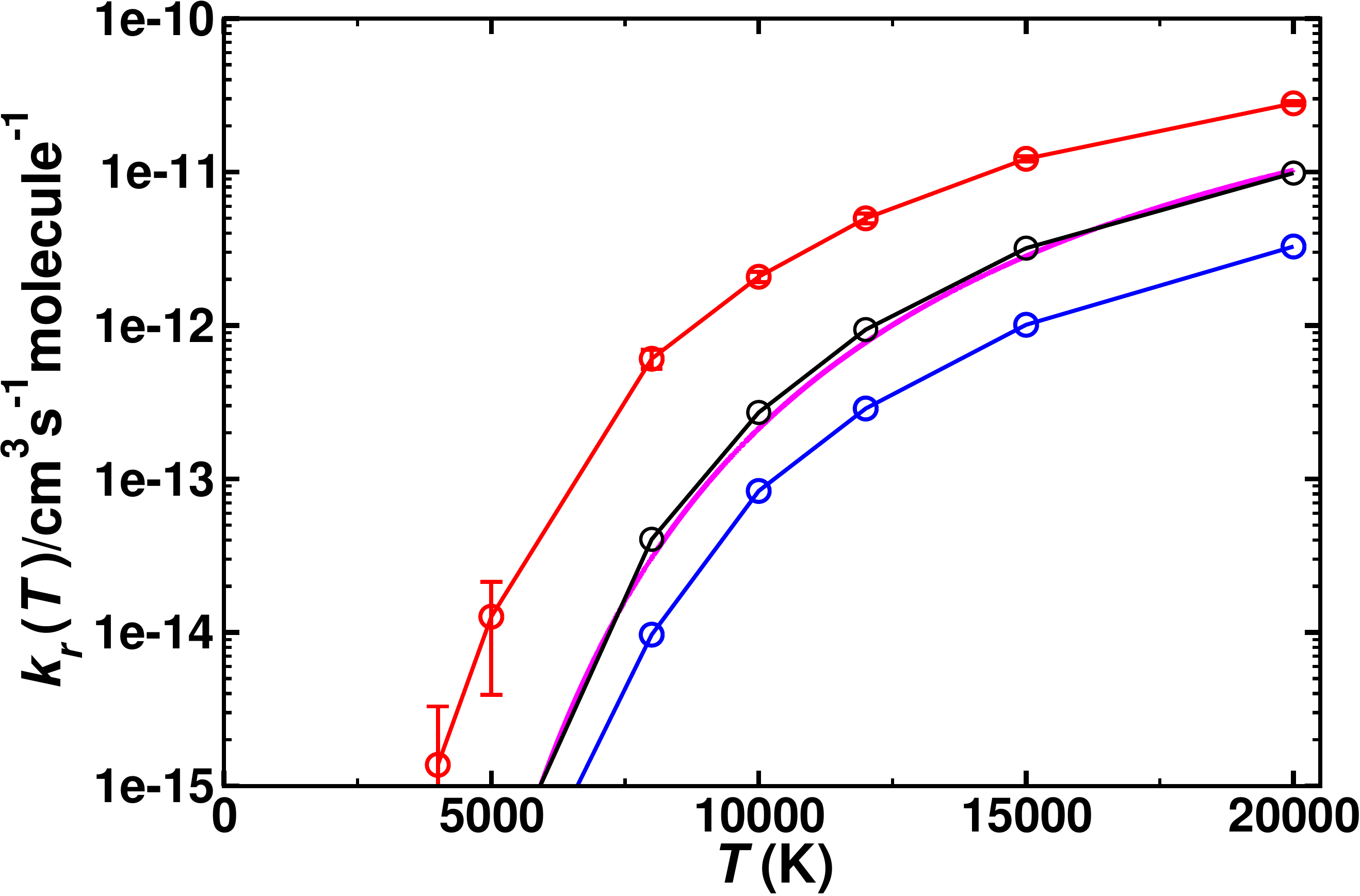}
    \caption{Thermal rate for the reverse ($k_{\rm r}$) reaction
      CO($^{1}\Sigma^{+}$)+ O($^{1}$D) $\rightarrow$ C($^3$P) +
      O$_2$($^3\Sigma_{g}^{-}$). The sum of the contributions of the
      singlet (red circles, with error bars from bootstrapping) and
      triplet (blue circles) states and their Boltzmann-weighted sum
      (black circles). The temperature range is from 5000-20000
      K. Comparison with recent theoretical work\cite{schwenke2016}
      (magenta solid line).}
\label{fig:ratereverse}
\end{center}
\end{figure}

\noindent
{\it For the reverse reaction,} CO($^{1}\Sigma^{+}$)+
O($^{1}$D)/O($^{3}$P) $\rightarrow$ C($^3$P) +
O$_2$($^3\Sigma_{g}^{-}$), similar simulations were carried out. As
this is an uphill process (Figure \ref{fig:fig1}), this channel only
opens at higher temperature, see Figure \ref{fig:ratereverse}. The
dynamics for CO($^{1}\Sigma^{+}$)+ O($^{1}$D) $\rightarrow$ C($^3$P) +
O$_2$($^3\Sigma_{g}^{-}$) involves the $^1$A$'$, $(2)^1$A$'$, and
$^1$A$''$ states (for numerical values see Table
S3), whereas that for CO($^1\Sigma^{+}$)+
O($^3$P) $\rightarrow$ C($^3$P) + O$_2$($^3\Sigma_{g}^{-}$) is
related to the $^3$A$'$ and $^3$A$''$ states, given in Table
S4. Compared with the forward rates, those for
the reverse reaction are typically 1 to 5 orders of magnitude
smaller. The reverse rates starting from O($^1$D) are larger by 1 to 2
orders of magnitude at high $T$ than those from O($^3$P) which is
consistent with the Boltzmann-weighted energy difference for the two
asymptotes.\\

\begin{tiny}
\begin{table}[h]
\begin{tabular}{c|c|c|r}
\hline
Forward  & $A$ & $n$ & $\epsilon$ \\
\hline 
C($^{3}$P) + O$_2$($^3 \Sigma^-_g$) $\rightarrow$ CO($^{1} \Sigma^{+}$)+ O($^1$D)  & \ \ 4.12 $\times10^{-12}$  &  0.45  &  2209 \\
\hline 
$^1\rm{A}'$  &  2.42$\times10^{-12}$ &  0.40 &  116 \\
$(2)^1\rm{A}'$  &  1.21$\times10^{-12}$ &  0.47 &   4506 \\
$^1\rm{A}''$  &  1.06$\times10^{-11}$ &  0.27 &  6639 \\
\hline
C($^{3}$P) + O$_2$($^3 \Sigma^-_g$) $\rightarrow$ CO($^{1} \Sigma^{+}$)+ O($^3$P)  & 3.50 $\times10^{-11}$ & 0.22  & 3513 \\
\hline
$^3\rm{A}'$ &   1.60$\times10^{-11}$ &  0.22 &   1891 \\
$^3\rm{A}''$ &  5.49$\times10^{-11}$ &  0.11 &  6789 \\
\hline
Total  &  1.56 $\times10^{-11}$  &  0.30  &  3018  \\
\hline 
\hline
Reverse  & \ \  \ \  & \ \  \ \ & \ \  \ \ \\
\hline
O($^{1}$D) + CO($^1 \Sigma^+$) $\rightarrow$ C($^3$P) +O$_{2}$($^3\Sigma_g^-$)  & 1.15 $\times10^{-10}$ & 0.11 &  49965  \\
\hline 
$^1\rm{A}'$ &  1.25$\times10^{-12}$ &  0.42 &  42273 \\
$(2)^1\rm{A}'$ &  4.60$\times10^{-12}$ &  0.32 & 50111 \\
$^1\rm{A}''$  &  5.28$\times10^{-13}$ &  0.53 & 46836 \\
\hline
O($^{3}$P) + CO($^1 \Sigma^+$) $\rightarrow$ C($^3$P) +O$_{2}$($^3\Sigma_g^-$)  & 1.52 $\times10^{-12}$ & 0.50  & 68903 \\
\hline
$^3\rm{A}'$ &   8.92$\times10^{-14}$ &  0.70 &   64167 \\
$^3\rm{A}''$ &  7.80$\times10^{-09}$ &  --0.37 &  83013 \\
\hline
Total  &  1.55 $\times10^{-10}$  &  0.09  &  71735  \\
\hline
\hline
\end{tabular}
\caption{Modified Arrhenius 3-parameter model for the forward
  C($^{3}$P) + O$_2$($^3 \Sigma^-_g$) $\rightarrow$ CO($^{1}
  \Sigma^{+}$)+ O($^1$D)/O($^3$P) and reverse O($^1$D)/O($^{3}$P) +
  CO($^{1} \Sigma^{+}$) $\rightarrow$ C($^3$P)
  +O$_{2}$($^3\Sigma_g^-$) reaction. $A$ in units of cm$^3$ s$^{-1}$
  molecule$^{-1}$ and $\epsilon$ in Kelvin. The temperature range for
  the modified Arrhenius fit is 5000 to 20000 K.}
\label{tab:fit}
\end{table}
\end{tiny}

\noindent
Table \ref{tab:fit} summarizes the parameters from fitting the raw
data to a modified Arrhenius expression $k(T)=AT^{n}
\exp{(-\epsilon/T)}$ for the forward and reverse processes for all
five PESs. It is noted that all forward processes involve a
comparatively small activation energy $\epsilon$ of a few hundred to a
few thousand Kelvin. All reverse rates have activation energies that
are at least one order of magnitude larger. The number of trajectories
that contribute to these rates varies from less than 1 \% to 55
\%. For the slowest process, the reverse reaction on the $^3$A$'$ and
$^3$A$''$ PESs originating from O($^3$P), at least an additional $5
\times 10^5$ trajectories were run at each temperature between 3000 K
and 20000 K and close to $10^6$ trajectories for $T \leq 1000$ K.\\

\begin{figure}[h]
    \begin{center}
      \includegraphics[scale=0.6]{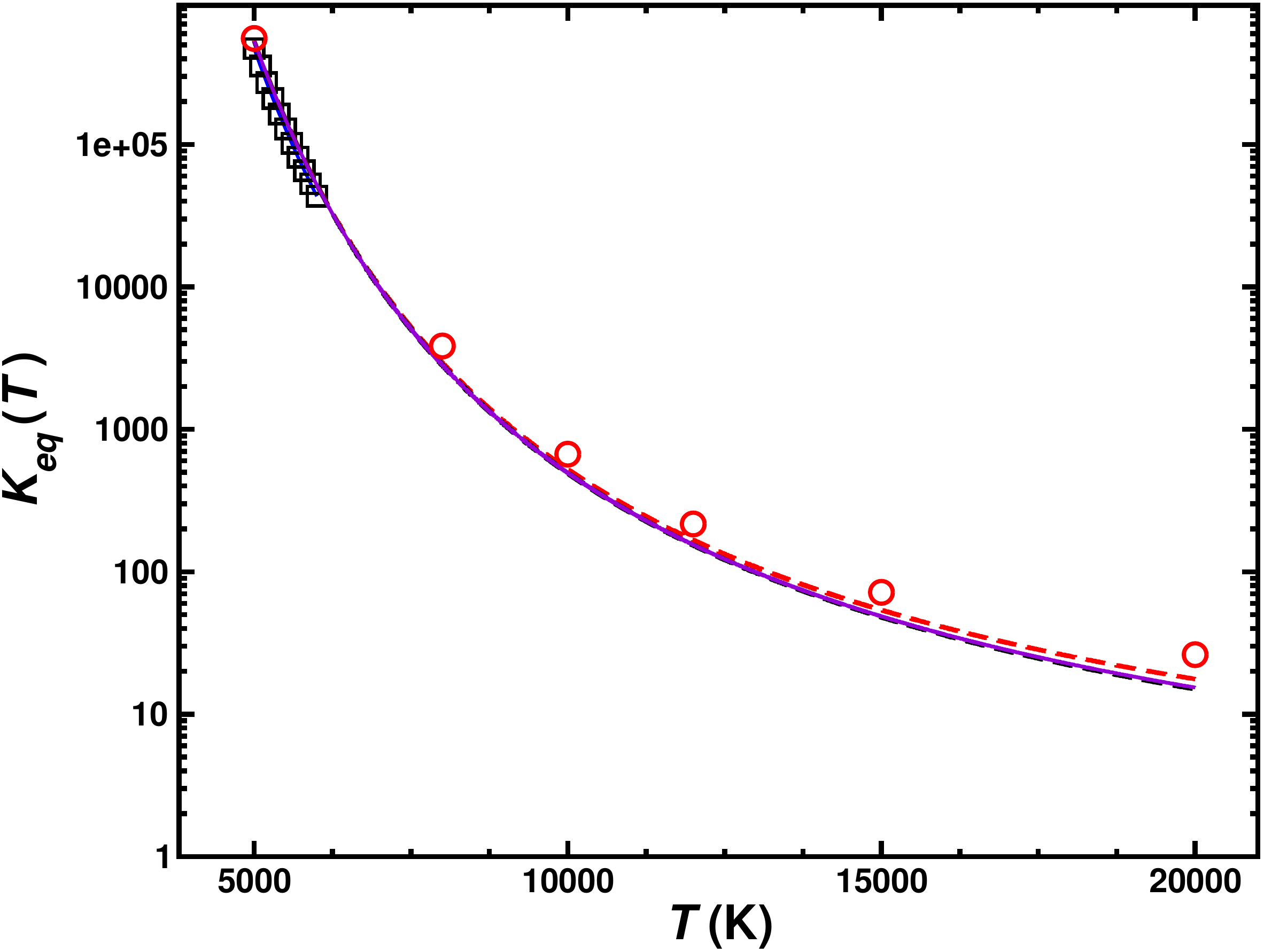}
    \caption{Equilibrium constant for the C($^3$P) +
      O$_2$($^3\Sigma_{g}^{-}$) $\leftrightarrow$
      CO($^{1}\Sigma^{+}$) + O($^1$D$_{1}$) ($k_{1}$) and C($^3$P) +
      O$_2$($^3\Sigma_{g}^{-}$) $\leftrightarrow$
      CO($^{1}\Sigma^{+}$) + O($^3$P) ($k_{2}$) reactions. The
      results from the JANNAF tables\cite{janaf} (black open squares),
      those derived from equilibrium statistical mechanics ($k_{1}$
      (red open circles and red dashed line), $k_{2}$ (black dashed
      line) and their Boltzmann-weighted total $w k_1 + (1-w) k_2$
      (purple solid line)) with those from the QCT simulations are
      compared.}
\label{fig:keq}
\end{center}
\end{figure}

\noindent
From the forward and reverse rates the equilibrium constant $K_{\rm
  eq}(T)$ can also be determined, see Figure \ref{fig:keq}. This
equilibrium constant was determined from the total forward and reverse
fluxes of the weighted sum of the singlet and triplet pathways
according to the data summarized in Table \ref{tab:fit}. Error bars
for the individual rates have been determined from bootstrapping and
are compared with results determined from statistical mechanics. The
equilibrium constant is only reported for temperatures 5000 K and
higher as the reverse reaction only opens at these temperatures, see
Figure \ref{fig:ratereverse}.\\

\noindent
A final process considered is the atom exchange reaction CO$_{\rm
  A}$($^{1}\Sigma^{+}$)+ O$_{\rm B}$($^{3}$P) $\rightarrow$ CO$_{\rm
  B}$($^{1}\Sigma^{+}$)+ O$_{\rm A}$($^{3}$P). For this process, on
the $^3$A$'$ state, rates ranging from $5 \times 10^{-16}$ cm$^{3}$
molecule$^{-1}$ s$^{-1}$ to $6 \times 10^{-11}$ cm$^{3}$
molecule$^{-1}$ s$^{-1}$ between 500 K and 20000 K were found, see
Table S5 and Figure S6. The rate
increases monotonically from values $\sim 10^{-16}$, consistent with
those measured experimentally,\cite{Jaffe1966} as a function of $T$
and is smaller than the measurement at 1820 K.\cite{Garnett1969} This
experimental value was an indirect measurement that required the
decomposition rate for N$_2$O and is presented without derived error
bars. The barrier for the atom exchange reaction inferred from the
low-temperature experiments is 6.9 kcal/mol (0.299 eV), which is also
what is found from the present work (Figures S7
and S8).\\
    
\noindent
A summary of all forward and reverse rates is provided in Figure
S5, and Tables S1 to
S5 report all numerical values for the temperature
dependent rates.\\

\subsection{Vibrational Relaxation}
\begin{figure}[h]
    \includegraphics[scale=0.5]{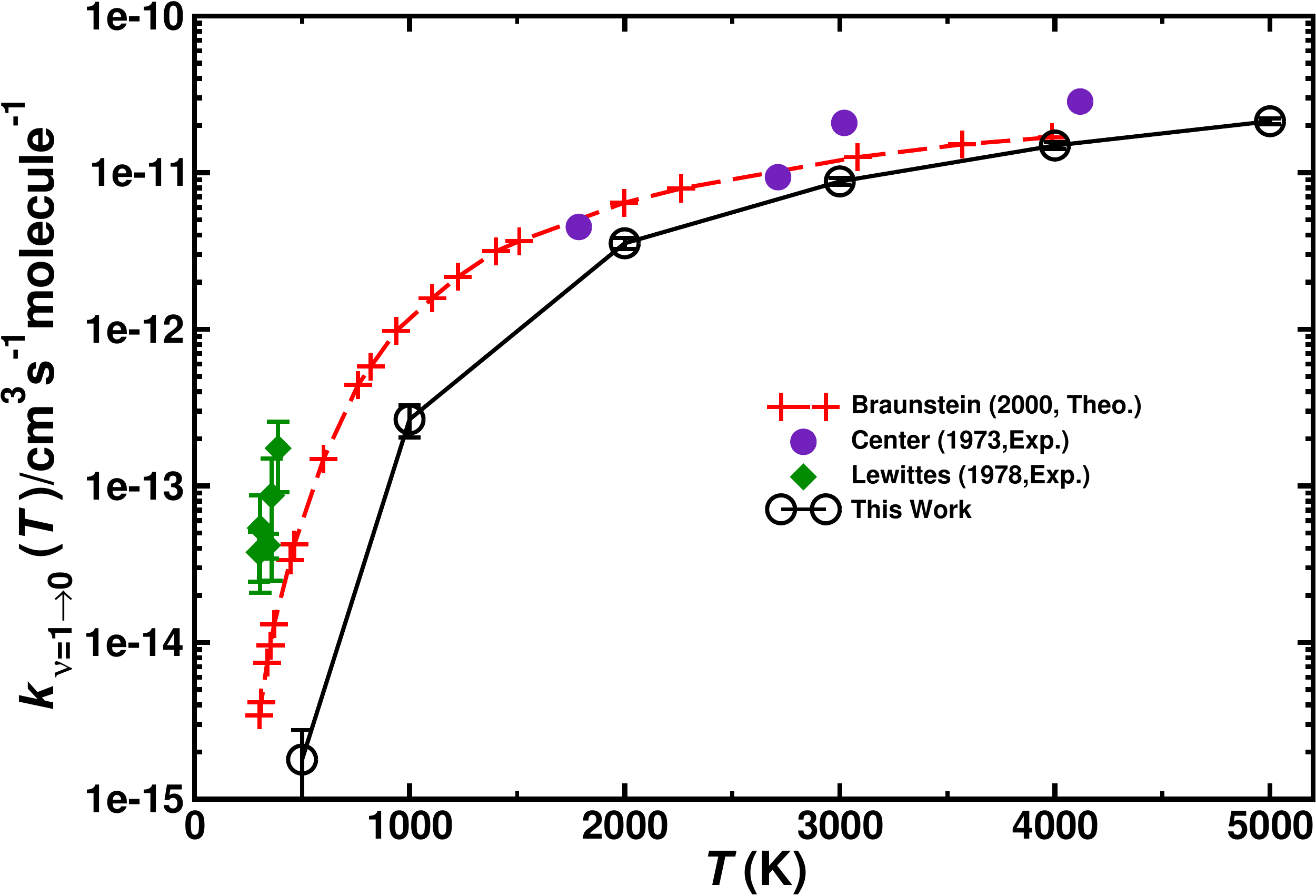}
    \caption{Total vibrational relaxation rate for O+CO($\nu=1$)
      $\rightarrow$ O+CO($\nu=0$). Total contribution
      ($^3$A$'$+$^3$A$''$) (closed black circles) ($g(e)=1/3$).
      Literature values are the symbols as
      indicated.\cite{Lewittes1978,Center1973,Braunstein2000}. $5
      \times 10^5$ trajectories were run at every temperature.}
\label{fig:vr}
\end{figure}

Vibrational relaxation (VR) of CO in its $v=1$ and $v=2$ states was
investigated for both, the singlet and triplet manifolds
separately. VR was investigated by running $5 \times 10^5$
trajectories at each temperature, ranging from 300 K to 5000 K, see
Table S6. The final vibrational state was determined
using Gaussian binning (GB) which has been shown to yield similar
results as histogram binning.\cite{bon97:183,bon04:106,konthesis}
Figure \ref{fig:vr} compares the individual and total VR rates with
those measured experimentally and Table \ref{tab:vr} reports the
rates. The computed rates are consistently lower than those from
experiments at lower temperatures. For $T > 2000$ K the rates are in
good agreement with experiments, though. In order to verify that the
underestimation is not due to neglect of higher electronically excited
states, the $(2)^3$A$''$ PES was also determined. This PES (not shown)
is mainly repulsive. Therefore, the VR rates for this state only
contribute $\sim 10$ \% of the rates for the $^3$A$'$ and $^3$A$''$
states at the highest temperatures. Hence, the differences between
experiment and simulations at lower temperatures are not due to
neglect of contributions from higher-lying electronic states.\\

\begin{table}[h]
\begin{tabular}{c|lllllll}
  & 500 K & 1000 K & 2000 K & 3000 K & 4000 K & 5000 K \\
  \hline
  $^3$A$'$ & 0.01 & 1.51 & 18.61 & 43.62 & 74.44 & 101.95 \\
  $^3$A$''$ & 0.01 & 1.08 & 14.90 & 39.70 & 67.00 & 99.30 \\
  $(2)^3$A$''$ & 0.00 & 0.07 & 1.90 & 4.71 & 7.83 & 11.73 \\
  \hline
  Total & 0.02 & 2.66 & 35.41 & 88.03 & 149.27 & 212.98 \\ \hline
\end{tabular}
\caption{Vibrational relaxation rates (in units of $10^{13}$\rateunit)
  $k_{\nu \rightarrow \nu'}$ for the collision of O($^3$P) with CO($^1
  \Sigma_{\rm g}$): O+CO($\nu=1$) $\rightarrow$ O+CO($\nu'=0$) for the
  $^{3}$A$'$, $^{3}$A$''$, and $(2)^3$A$''$ states and the total
  contribution using GB.}
\label{tab:vr}
\end{table}

\begin{figure}[h!]
    \includegraphics[scale=0.75]{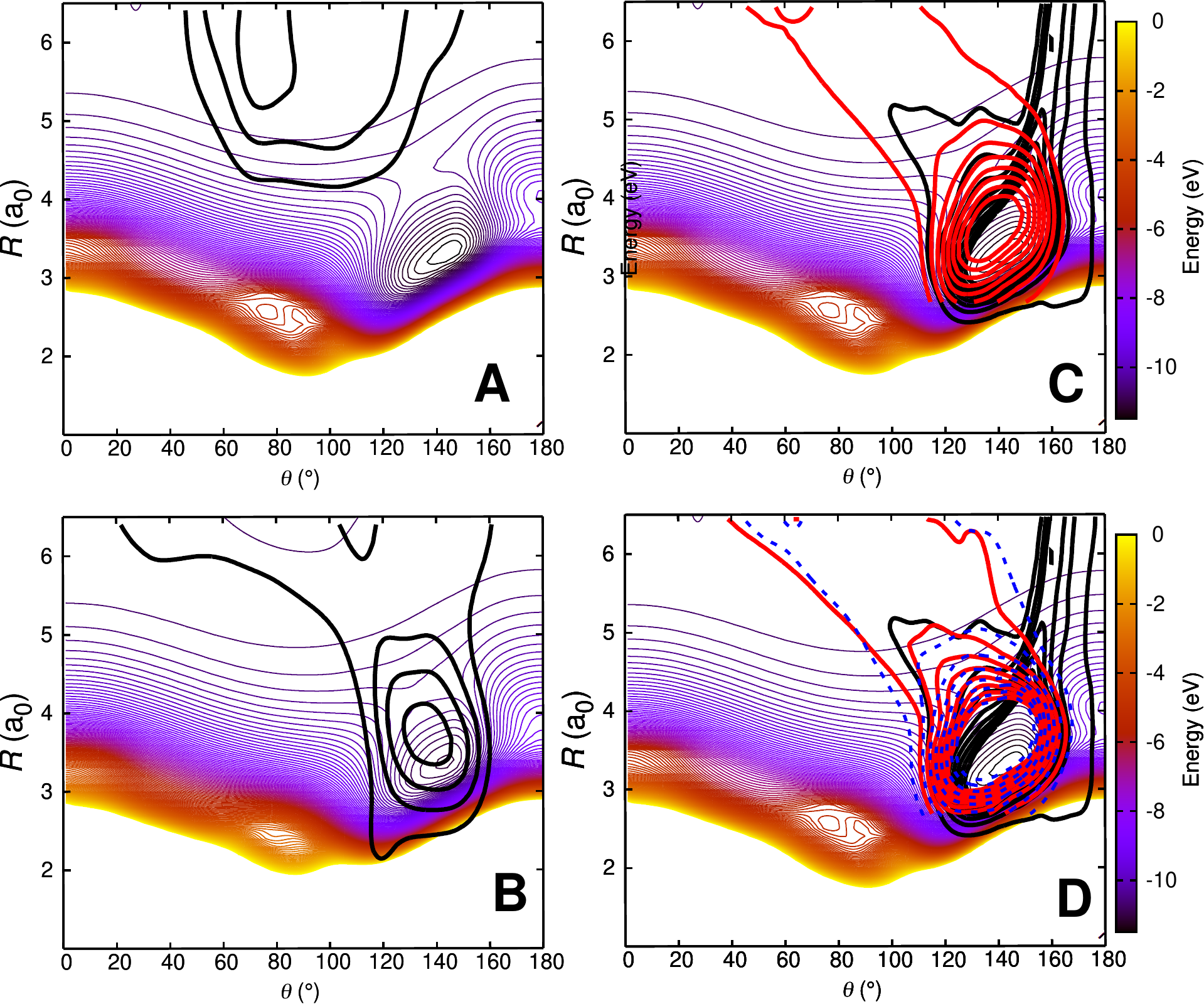}
    \caption{Density map for O+CO($\nu=1$) collisions at 1000 K on the
      relaxed $^3$A$'$ PES. Panel A: O+CO($\nu=1$) $\rightarrow$
      O+CO($\nu'=1$); Panel B: O+CO($\nu=1$) $\rightarrow$
      O+CO($\nu'=0$); Panel C: O$_{\rm A}$+CO$_{\rm B}$($\nu=1$)
      $\rightarrow$ O$_{\rm B}$+CO$_{\rm A}$($\nu'=1$) and Panel D:
      O$_{\rm A}$+CO$_{\rm B}$($\nu=1$) $\rightarrow$ O$_{\rm
        B}$+CO$_{\rm A}$($\nu'=0$). For the reactive trajectories
      (panels C and D), two coordinate systems are used: one for the
      reactant (black density) in which the CO$_{\rm B}$ diatom is the
      distance $r$ and the separation of atom O$_{\rm A}$ from the
      center of mass is the distance $R$; the second coordinate system
      is for the product state (red density) for which the CO$_{\rm
        A}$ diatom is the distance $r'$ and the separation of atom
      O$_{\rm B}$ from the center of mass is the distance $R'$. The
      dashed blue isocontours in panel D are for O$_{\rm A}$+CO$_{\rm
        B}$($\nu=1$) $\rightarrow$ O$_{\rm B}$+CO$_{\rm
        A}$($\nu'=2$). The density map for the trajectories is
      superimposed on a relaxed 2D RKHS PES where $2.00<r<2.30$
      a$_{0}$ (turning points). For all density maps 1500 trajectories
      were used to generate the 2d densities which were smoothed using
      kernel density estimation (KDE) as implemented in the R software
      package\cite{Rcore}.}
\label{fig:dmap}
\end{figure}

\noindent
In order to better characterize to which parts of the PESs the
individual processes are sensitive to, density maps were determined as
follows. For each initial condition a trajectory can be attributed to
one of the 4 possible outcomes: a) no vibrational relaxation, no
reaction: O+CO($\nu=1$) $\rightarrow$ O+CO($\nu'=1$) b) vibrational
relaxation without reaction: O+CO($\nu=1$) $\rightarrow$
O+CO($\nu'=0$), c) no vibrational relaxation but with atom exchange:
O$_{\rm A}$+CO$_{\rm B}$($\nu=1$) $\rightarrow$ O$_{\rm B}$+CO$_{\rm
  A}$($\nu'=1$), and d) vibrational relaxation with atom exchange:
O$_{\rm A}$+CO$_{\rm B}$($\nu=1$) $\rightarrow$ O$_{\rm B}$+CO$_{\rm
  A}$($\nu'=0$). Then, all trajectories for a given class were
combined and a 2-dimensional histogram was generated and smoothed
using kernel density estimation (KDE).\cite{parzen:1962} The resulting
2-dimensional distribution was then projected onto the relaxed PES for
the corresponding state, see Figure \ref{fig:dmap}.\\

\begin{table}[h!]
  \caption{Average contact time ($\tau_{c}$ in fs, for definition see
    text) and number $N$ of trajectories for each final state for
    $N_{\rm tot} = 5 \times 10^5$ trajectories for each of the
    processes considered. In each case the difference $(N_{\rm tot}-
    N)$ are fly-by trajectories. This table reports the cases $\nu=2
    \rightarrow \rm{x}$ and $\nu=1 \rightarrow \rm{x}$ for both
    reactive and non-reactive events.}
\begin{center}
\begin{tabular}{c|c|c|c||c|c}
  \hline
     &\multicolumn{2}{r}{relaxing} & & \multicolumn{2}{c}{nonrelaxing}\\
        \hline  
\hline
reacting &  $\nu=2 \rightarrow 1$   &  $\nu=2 \rightarrow 0$  &  $\nu=1 \rightarrow 0$  &  $\nu=2 \rightarrow 2$  &  $\nu=1 \rightarrow 1$ \\
    \hline
     $N$   & 300 & 230 & 681 & 340 & 745  \\
     \hline
     $\tau_{c}$ & 210 & 207 & 333 & 241 & 301  \\
    \hline
    \hline
non-reacting & \multicolumn{2}{r}{} & & \multicolumn{2}{c}{}\\
    \hline
    $N$ & 480 & 440 & 1579 & 65790  & 117367  \\
    \hline
    $\tau_{c}$  & 67 & 155 & 156 & 33 & 36\\
    \hline
    \hline
  \end{tabular}
    \end{center}
\label{tab:vr2}
\end{table}

\noindent
Panel \ref{fig:dmap}A shows that nonrelaxing trajectories sample
regions in the long range without penetrating into the strongly
interacting region around $(R=3.2 \rm{a}_0, \theta=
150^\circ)$. Contrary to that, nonreactive, relaxing trajectories of
the type O+CO($\nu=1$) $\rightarrow$ O+CO($\nu'=0$) access the
strongly interacting region and sample it before leaving this region
again, see Figure \ref{fig:dmap}B. For the reactive trajectories
(O$_{\rm A}$+CO$_{\rm B}$ $\rightarrow$ O$_{\rm B}$+CO$_{\rm A}$), see
Figures \ref{fig:dmap}C and D, all trajectories enter the strongly
interacting region along $\theta \sim 160^\circ$ (black density).\\

\noindent
After the reaction, the product (CO$_{\rm A}$) can either remain
vibrationally excited (Figure \ref{fig:dmap}C; no relaxation), or its
vibrational state can change (CO$_{\rm A}$($v'=0$) or CO$_{\rm
  A}$($v'=2$)). The highest vibrational state in the products after
reaction in these trajectories (run at 1000 K) is $v' = 3$. The
probability distributions of the products from reactive collisions in
Figures \ref{fig:dmap}C and D are in red (for $v' =0$, relaxation) and
in blue (for $v' = 2$, further excitation). The shape of the red and
blue probability distributions in Figures \ref{fig:dmap}C and D can
already be anticipated from the relaxed PES for the CO+O channel, see
Figure S9. Starting from around the minimum on the
PES at $(R = 3.2 a_0, \theta = 140^\circ)$ these densities follow the
path indicated by the green isocontour at --10.5 eV in Figure
S9 through the constriction indicated as a red
cross. A different perspective that could be taken is to refer to all
reactive trajectories as ``vibrationally relaxing'' because the quanta
initially present in CO$_{\rm B}$ are destroyed upon dissociation of
CO$_{\rm B}$. However, experimentally, the final states CO$_{\rm B}
(v'=1)$ and CO$_{\rm A} (v'=1)$ can not be distinguished. Hence
separation into 4 separate cases is meaningful in analyzing the
trajectories.\\

\begin{figure}
    \centering
    \includegraphics[scale=0.3]{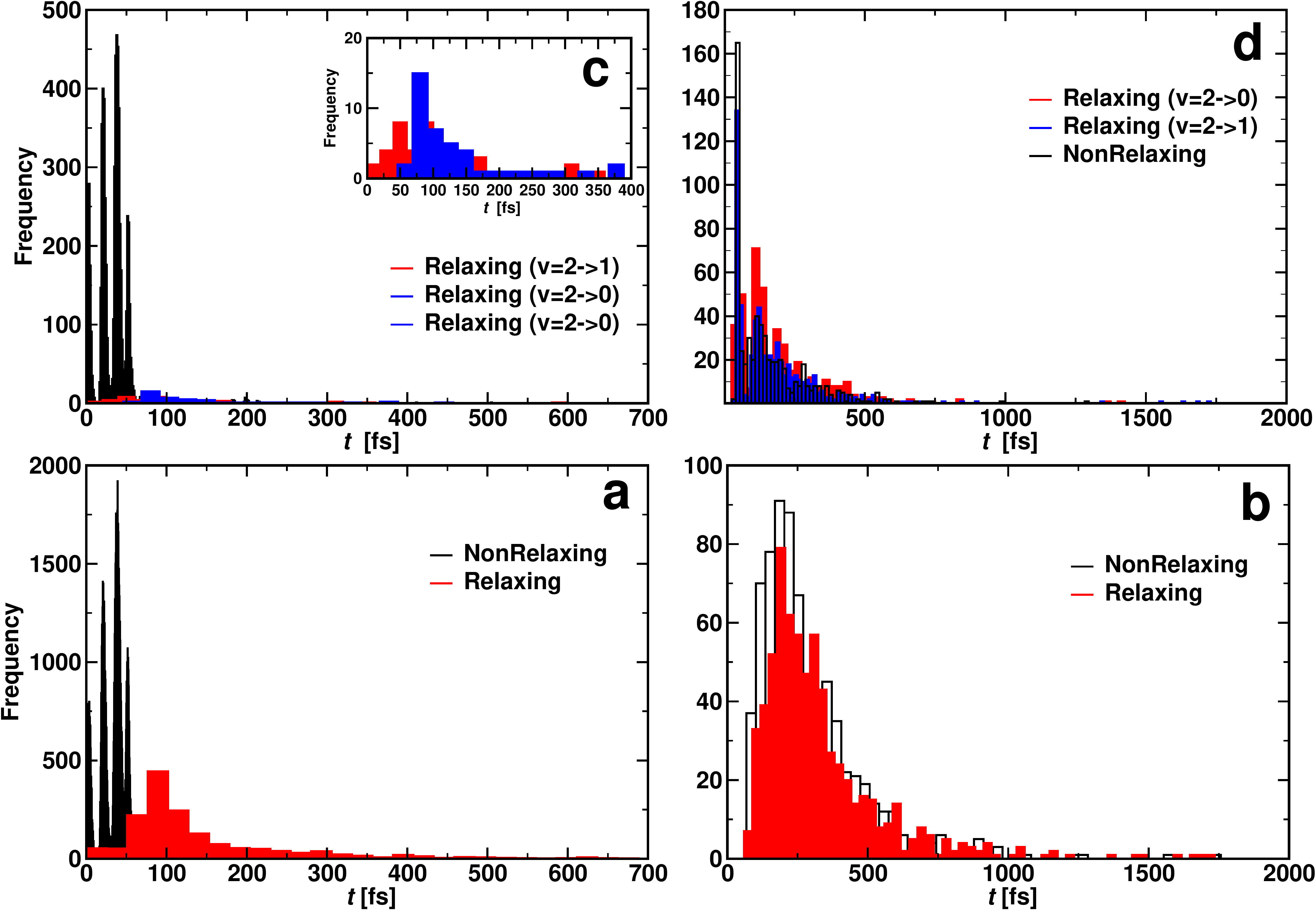}
    \caption{Contact time histogram for O$_{\rm A}$+CO$_{\rm B}$
      $\rightarrow$ O$_{\rm B}$+CO$_{\rm A}$ (reacting: panel a and c)
      and O$_{\rm A}$+CO$_{\rm B}$ $\rightarrow$ O$_{\rm A}$+CO$_{\rm
        B}$ (non reacting: panel b and d). Bottom panels for CO($v=1$)
      and top panel for CO($v=2$). Panel c inset illustrates that
      relaxing two quanta ($\nu = 2 \rightarrow 0$, blue distribution)
      takes longer than relaxing one quantum ($\nu = 2 \rightarrow 1$,
      red distribution). Additional analysis of the data from panel c
      is provided in Figure S10. Rates for the atom
      exchange reaction are given in Table S5.}
    \label{fig:lifetime}
\end{figure}

\noindent
It is also of interest to consider the distribution of contact times
$\tau_c$ for each of the scenarios. This quantity was taken as the
first instance along the trajectory for which the sum $\sigma$ of all
three atom distances is smaller than 12 a$_0$ in the entrance channel
until the point at which $\sigma > 12$ a$_0$ along the exit
channel. This was done for CO initially in its $v=1$ and $v=2$ states,
respectively. The average contact times are reported in Table
\ref{tab:vr2} and their distributions are shown in Figure
\ref{fig:lifetime}. It should, however, be noted that the average
$\tau_c$ only incompletely characterize the underlying distribution
$P(\tau_c)$ because the distributions are either structured (Figures
\ref{fig:lifetime}a and c) or extend to times more than 10 times the
most probable value as in Figures \ref{fig:lifetime}b and d.\\

\noindent
For reacting trajectories and non-reacting but relaxing trajectories,
the contact time $\tau_c$ decreases with increasing vibrational
excitation. This differs for non-reacting, relaxing
trajectories. Their average contact times appear to be determined by
the final vibrational state. For relaxation to $v' = 0$ the average
vibrational relaxation time is $\sim 150$ fs which shortens to $\sim
70$ fs for relaxation to $v' = 1$ with initial $v = 2$. This is in
contrast to the non-relaxing non-reacting trajectories which appear to
be independent of vibrational excitation.  The $\tau_c$ for these
trajectories is of the order of 30 fs which is roughly the minimum
time required for one collision.\\

\noindent
When considering the lifetime distributions it is found that those
involving reacting trajectories display a regular pattern of peaks,
see Figures \ref{fig:lifetime}a and c. The specific case for
relaxation from $(v=1) \rightarrow (v' =0)$ is shown in Figure
S10. It is noticeable that the probability to find
trajectories that react but do not relax $P(\tau_c)$ can be zero and
reaches maximal values for other values for the lifetime. Fourier
transformation of this signal yields frequencies between 1824
cm$^{-1}$ and 2529 cm$^{-1}$, see Figure S10. These
frequencies, which are in the range of typical CO stretch frequencies,
can be understood as ``gating modes'' that allow the reaction to
occur, similar to what was found for proton transfer in protonated
ammonia dimer.\cite{MM.n2h7:2002}\\

\noindent
It is also of interest to consider the geometries sampled for the
C+O$_{2} \longrightarrow$ CO+O($^1$D) reaction on the $^1$A$'$ PES
depending on the temperature from which the initial conditions were
generated. This was done for $T=15$ K and $T=10000$ K. For reactive
trajectories at low temperature the global minimum is extensively
sampled (see Figure S11A) whereas at high
temperature this region is not sampled at all as shown in Figure
S11B. Hence, collisions at different temperatures
are expected to sample complementary regions of the 3d PES.\\

\section{Discussion and Conclusions}
The present work reports thermal and vibrational relaxation rates from
QCT simulations on the five lowest PESs of the [COO]
systems. Comparison with experiment is favourable for thermal rates
and vibrational relaxation rates at high temperatures. For the atom
exchange rate, agreement is rather more qualitative, with an overall
offset in the energetics of 300 K (0.026 eV). Additional analyses are
carried out in the following to provide an understanding of remaining
disagreements between experiment and simulations.\\

\noindent
One interesting comparison can be made with state-to-state cross
section measurements for the C($^{3}$P) + O$_{2}$($^3\Sigma_{g}^{-}$)
$\leftrightarrow$ CO($^{1}\Sigma^{+}$)+ O($^{1}$D) reaction at small
collision energies.\cite{costes:1998} These experiments used a pulsed
nozzle through which the O$_2$ expanded into the vacuum. The O$_2$
internal state distribution was not measured directly but expected to
be very cold.\cite{costes:1998} Hence, it is likely that mostly
O$_2$($v = 0$) with low $j_{\rm max}$ was populated. Such experiments
found that excitation of CO($v'=16$) occurs for all collision energies
$E_{\rm c}$ whereas population of CO($v'=17$) is only possible with an
excess of $E_{\rm c} > 0.04$ eV. Using bound state energies for CO
derived from experiment\cite{chandra:1996} and accounting for the 0.04
eV required to open the CO$(v'=17)$ channel, the energy difference
between CO$(v=0)$ and CO$(v'=17)$ is 4.037 eV. Including zero point
energy for CO and O$_2$, the difference between the C($^3$P)+O$_2$ and
CO+O($^3$P) channels from experimental data is 4.075 eV. This differs
by 0.085 eV from the value at the MRCI level of theory which is 3.990
eV.\\

\noindent
From semiclassical calculations on the present PESs the CO($v'=17$)
state is at 4.140 eV. This compares with the difference in electronic
energies (3.990 eV) and differences in the CO and O$_2$ zero point
energies of 3.952 eV. Hence, $E_{\rm c} = (4.140-3.952) = 0.188 $ eV
is required to open the CO($v'=17$) channel. QCT simulations starting
from Boltzmann-distributed $(v,j)$ initial conditions find that the
population of the CO($v'=16$) decays exponentially with increasing
$E_{\rm c}$ (Figure S12 left panel) which is
consistent with experiments.\cite{costes:1998} Because expansion
through a nozzle does not necessarily yield Boltzmann-distributed
initial conditions and the experimental beam was deemed ``very
cold'',\cite{costes:1998} the final state distributions were also
separated into those originating from O$_2 (v=0)$ (open circles in
Figure S12) and those from O$_2 (v >0)$ (solid line
in Figure S12). For CO($v'=16$) all distributions
follow the same overall behaviour.\\

\noindent
Conversely, for CO($v'=17$) considering the final state distribution
from initial O$_2 (v=0)$ has an onset at $\sim 0.05$ eV (inset Figure
S12 right panel) with a dependence on $E_{\rm c}$
consistent with experiment\cite{costes:1998} whereas including all
initial $v-$states for O$_2$ and those starting from O$_2 (v >0)$
again show a decaying probability distribution with increasing $E_{\rm
  c}$. Because both, initial $v$ and $j$ are probably ``cold'', it is
meaningful to consider final CO($v'=17$) distributions originating
from different $j_{\rm max}$ values for the parent O$_2$
molecule. With decreasing $j_{\rm max}$ the CO($v'=17$) channel opens
with increasing values of $E_{\rm c}$. For $j_{\rm max}^{\rm O_2} <
30$, the onset occurs at 0.05 eV and shifts to $\sim 0.15$ eV for
$j_{\rm max}^{\rm O_2} < 10$, which is consistent with the estimate of
0.188 eV based entirely on energetic arguments above.  A temperature
of $T = 300$ K corresponds to O$_2 (j=12)$ but the corresponding
(nonequilibrium) distribution probably extends to higher
$j-$values. Hence an estimated onset of generating CO$(v'=17)$ for
$E_{\rm c} \in [0.05,0.10]$ eV is expected from the present
simulations. This corresponds to a difference of 0.01 eV to 0.06 eV
from experiment on a scale of 4 eV, which is an error of 1 \% at
most.\\

\noindent
For the deactivation of O($^1$D) to O($^3$P) in the atmosphere early
models performed well for the observed data available at that
time.\cite{Tully:1975} The CO case was categorized as one that is
dominated by the configuration of a critical region where a crossing
between the single PESs originating from the O($^1$D) channel cross
the triplet PESs leading to O($^3$P). For the crossing dynamics a
Landau-Zener model was assumed. This simple approach lead to a
predicted rate of $8.0 \times 10^{-11}$ \rateunit at 300K which was
within the error of experimental measurements of 7.6 and $7.3 \times
10^{-11}$ \rateunit.\cite{husain:1973.1,stedman:1973} Following this,
the deactivation of O($^1$D) by CO was measured and the rate obtained
was fit by the expression $(4.7 \pm 0.9) \times 10^{-11} \exp{((126
  \pm 33)/RT)}$ which yields a rate of $5.8 \times 10^{-11}$ \rateunit
at 300 K. Assuming $\sim 5 \times 10^{-10}$ \rateunit for the
collision rate, this implies a $\approx 10$\% efficiency for
deactivation of O($^1$D) to O($^3$P) at 300 K. Based on this low
efficiency the crossings between the singlet and triplet manifolds are
not expected to have a large impact on the formation, exchange or
relaxation of the reaction.\\

\noindent
As Tully pointed out, deactivation depends on the specific crossing
geometry of the PESs; in this case the singlet and triplet
surfaces. Figures S13 and S14 show
the crossings of the $^3$A with the $^1$A surfaces on PESs evaluated
at the inner (Figure S13) and outer (Figure S13) turning points for the CO($v=0$)
vibration. When starting from the CO$_{\rm A}$+O$_{\rm B}$ side of the
reaction, as was previously mentioned, Figure \ref{fig:dmap} shows
that the active reactions sample a channel near $140^\circ$ that
brings the outgoing O$_{\rm A}$ atom into approximately $R = 3.5$
a$_0$. At low temperature, starting from CO+O($^3$P), it would be
possible to cross from any of the $^3$A surfaces onto the $^1$A$'$
surface to lead to ground state CO$_2$. However, in a collisionless
environment the complex will still have sufficient energy to return to
the $^1$A$'$ PES and will have to cross with a $^3$A surface to leave
as O($^3$P). This may affect vibrational energy transfer or the
exchange reaction and may be the reason for the shifts in the onset
seen between the experiment and QCT such as in Figure \ref{fig:vr} at
low temperature. Starting from CO+O($^1$D) and traveling along the
$^1$A$'$ surface crosses all $^3$A surfaces while the (2)$^1$A$'$ and
$^1$A$''$ surfaces only cross the repulsive (2)$^3$A$''$ surface. At
temperatures lower than that required to form C($^3$P)+O$_2$ these
trajectories can potentially cross on to the $^3$A surfaces and then
return to the CO+O($^3$P) state although it would be at high CO
vibrational state.\\

\noindent
One finding of the present work is the role ``gating'' plays in the
different processes considered here. For one, vibrational relaxation
with atom exchange displays gating in the contact time distributions
which hints at a time-dependent barrier in the [COO] collision
complex. This is explicitly seen in the barriers for the CO$_{\rm
  A}$($^{1}\Sigma^{+}$) + O$_{\rm B}$($^{3}$P) $\rightarrow$ CO$_{\rm
  B}$($^{1}\Sigma^{+}$) + O$_{\rm A}$($^{3}$P) atom exchange reaction
on the $^3$A$'$ PES (Figure S8). Depending on the
phase of the CO vibration at which the impinging oxygen atom collides
with the diatomic molecule, the barrier for formation of the collision
complex is either high or low. Such processes are particularly
susceptible to zero-point vibrational effects which can not be
captured in QCT simulations. Specifically, the vibrational
wavefunction does not produce the same spatial probability
distribution at low $v$ as the classical trajectory.  This results in
differences in sampling times for when the gate is open versus
closed. The rates from QCT simulations should, therefore,
underestimate the true rates, in particular at low temperatures. This
is indeed found for vibrational relaxation, see Figure \ref{fig:vr}
and for the atom exchange reaction S6. As the
vibrational relaxation rates include both, processes with and without
atom exchange, and the CO vibration-dependent barriers only affect
trajectories with atom exchange, it is conceivable that vibrational
relaxation without atom exchange is not affected by these effects.\\

\noindent
Including zero-point effects is likely to improve the comparison
between calculations and experiments. Furthermore, nonadiabatic
effects may further improve comparison with experiment, in particular
for the processes leading from CO$_2$ to the O+CO asymptotes. Analysis
of vibrational relaxation demonstrates that depending on the process
considered (with or without reaction), different parts of the
fully-dimensional PES are sampled. This is also true for reactions at
low (15 K) and higher (1000 K) temperatures, respectively. Together
with suitable information from experiment the underlying PESs could be
further improved from techniques such as
morphing\cite{MM.nehf:1999,bowman:1991} or Bayesian
inference.\cite{panesi:2020}\\

\noindent
In conclusion, the present work provides a comprehensive
characterization of the energetics and dynamics of the reactive [COO]
system involving the lowest five electronic states. Many findings
provide good agreement between simulations and experiments but it is
also found that disagreements can be traced back to neglecting quantum
mechanical effects at low temperatures. Additional experiments for
this important system will provide a more complete understanding of
the reactions involving both asymptotes.

\section*{Acknowledgments}
This work was supported by the Swiss National Science Foundation
through grants 200021-117810, and the NCCR MUST.

\bibliography{ref2}

\end{document}


\date{\today}

\begin{figure}[h!]
  \begin{center}
    \includegraphics[scale=0.4]{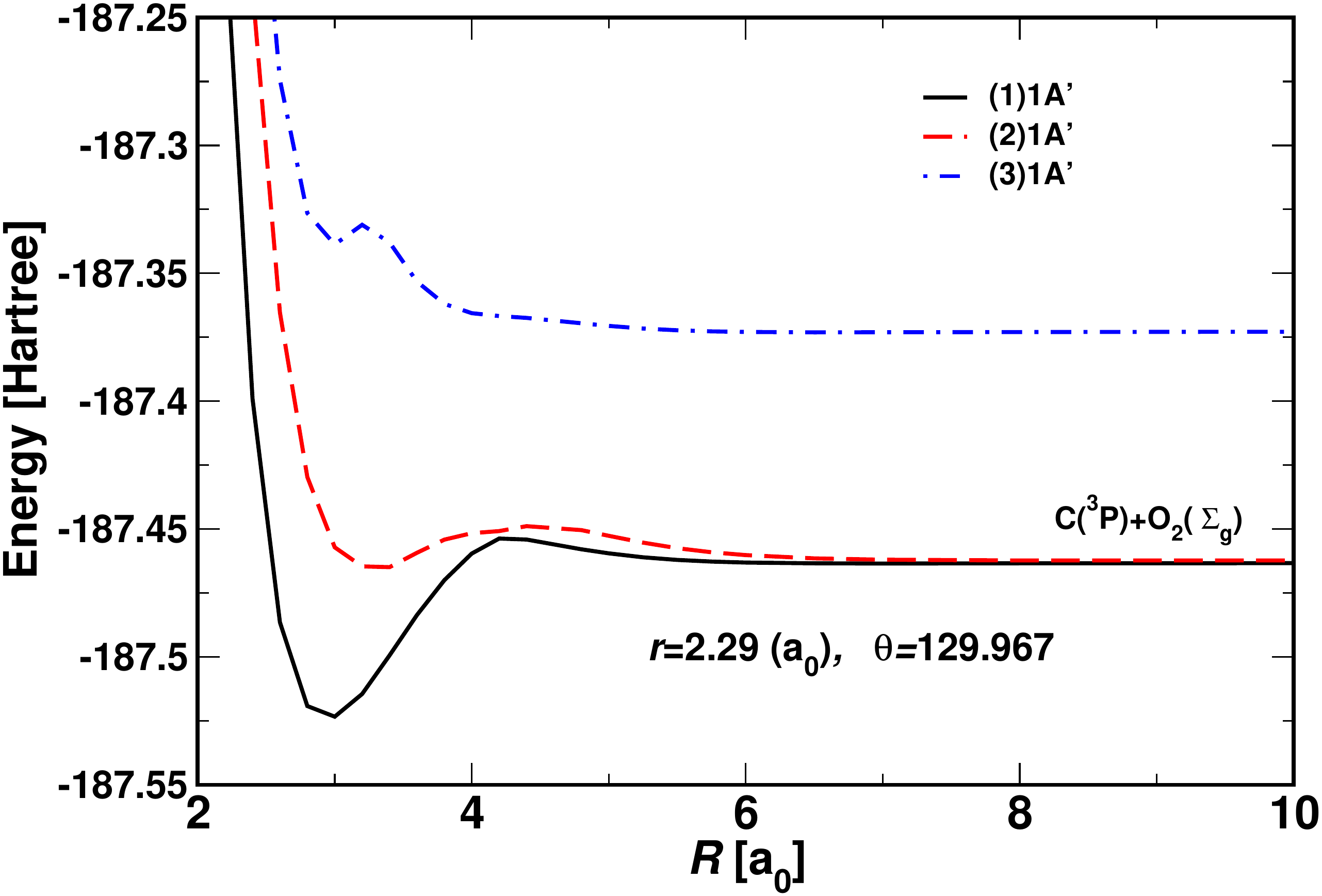}[b]
    \caption{Dissociation curves for C($^{3}$P) + O$_{2}$($^{3}
      \Sigma_{g}$) at $r_{\rm O-O} = 2.29$ a$_{0}$ and $\theta =
      129.95 ^{\circ}$ with varying $R$. Calculations at SA-CASSCF
      level with 3 considered states of the singlet $^{1}$A$'$ state.}
\label{sifig:casscf}
\end{center}
\end{figure}

\begin{figure}
  \begin{center}
    \includegraphics[scale=1.3]{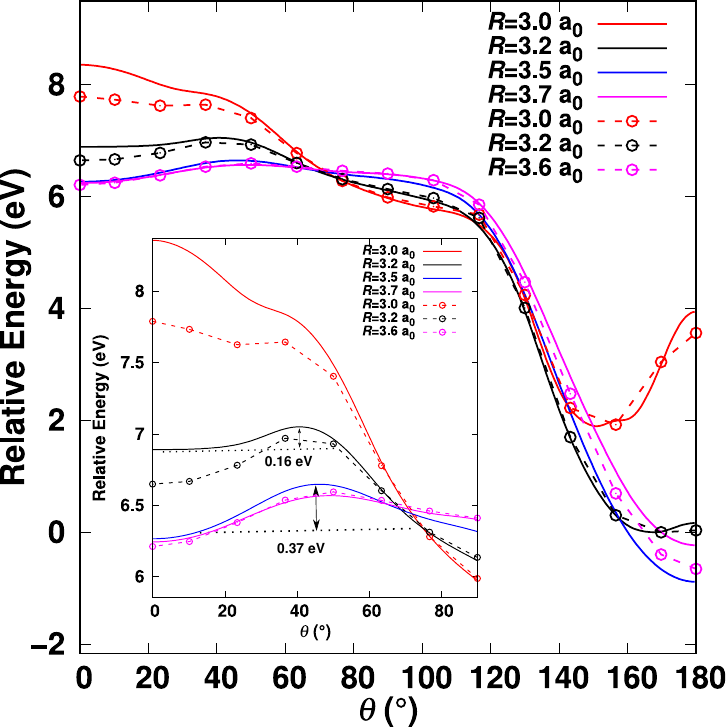}
\caption{Angular cuts through the $^1$A$'$ MRCI-PES (red) and CCSD(T)
  (black) for different values of $R$. At both levels of theory the
  COO structure is found to be a local minimum, stabilized by $\sim 9$
  (MRCI) and $\sim 5$ kcal/mol (0.39 and 0.22 eV) (CCSD(T)),
  respectively. The energy of the COO conformation lies 170 kcal/mol
  (7.37 eV) above the global minimum. The zero of energy is the
  minimum for $R=3.2$ a$_{0}$ for both the MRCI-RKHS (solid lines) and
  \textit{ab-initio} CCSD(T) (open circles and dashed lines), using
  the aug-cc-pVTZ basis set in both calculations.}
\label{sifig:coo}
\end{center}
\end{figure}

\begin{figure}
\includegraphics[scale=0.6]{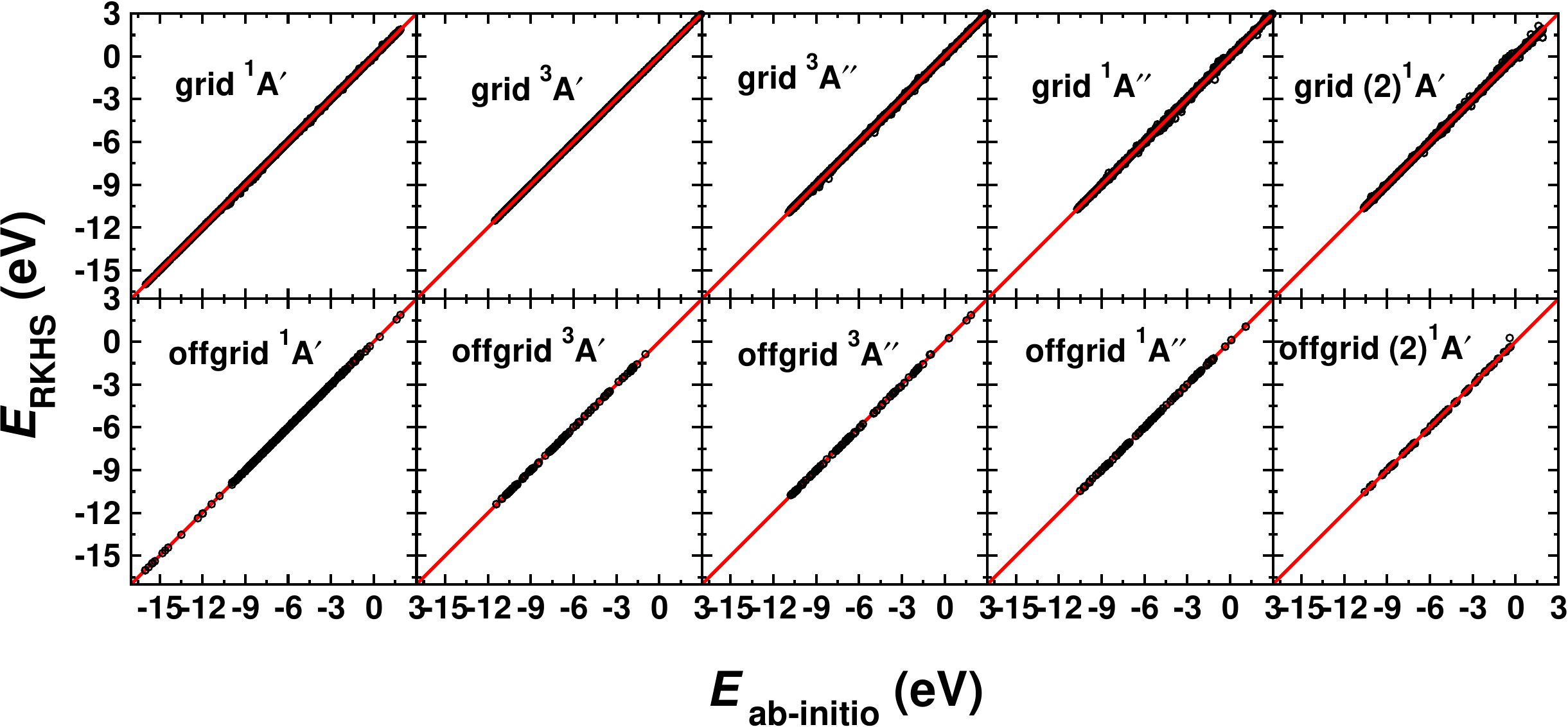}
\caption{Correlation between MRCI/aug-cc-PVTZ ($x-$axis) and RKHS
  energies ($y-$axis) for 23230 ($^1$A$'$), 9114 ($^3$A$'$), 20076
  ($^3$A$''$), 19560 ($^1$A$''$), and 20696 ($(2)^1$A$'$) for grid
  points and 618, 201, 194, 188 and 134 offgrid points for the
  $^1$A$'$,$^3$A$'$, $^3$A$''$, $^1$A$''$, and $(2)^1$A$'$ surfaces,
  respectively. The zero of energy is the O+O+C dissociation limit.
  The $R^2$ value for the grid points are (0.99998, 0.99999,
  0.99996,0.99984, 0.99984) and for off-grid points (0.99993, 0.99991,
  0.99990,0.99991, 0.99941) for the ($^1$A$'$,$^3$A$'$,
  $^3$A$''$,$^1$A$''$) surfaces, respectively. The corresponding root
  mean squared errors (RMSE) for the $^1$A$'$,$^3$A$'$, $^3$A$''$,
  $^1$A$''$ and $(2)^1$A$'$ surfaces are (0.01, 0.01, 0.02, 0.05,
  0.04) eV (0.33, 0.27, 0.56, 1.04, 0.81) kcal/mol for the grid points
  and (0.03, 0.03, 0.04, 0.03, 0.03) eV (0.65, 0.72, 0.89, 0.59, 0.75)
  kcal/mol for offgrid points.}
\label{sifig:fig1}
\end{figure}

\begin{figure}
    \centering \includegraphics[scale=0.5]{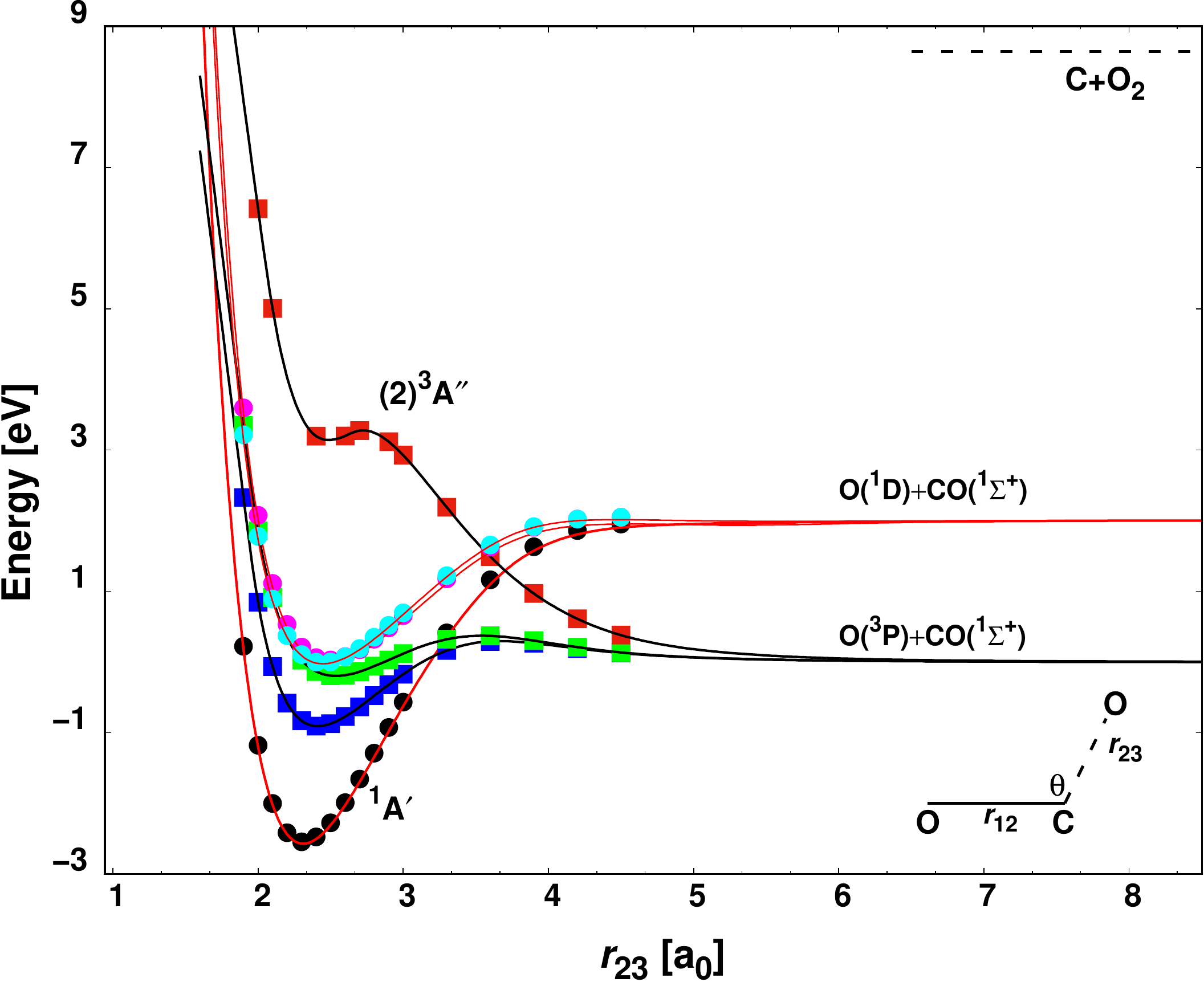}
    \caption{Ab initio calculated (points) and RKHS energies (solid
      lines) at a fixed CO bond length 1.2 \AA\/ (2.267 eV)and OCO
      angle $\theta=120.0^{\circ}$ for the $^1$A$'$ (black solid
      circle), 2$^1$A$'$ (cyan solid circle), $^1$A$''$ (magenta solid
      circle), $^3$A$'$ (blue solid square), $^3$A$''$ (green solid
      square), 2$^3$A$''$ (red solid square).}
\label{sifig:fig4}
\end{figure}

\begin{figure}
\includegraphics[scale=0.35]{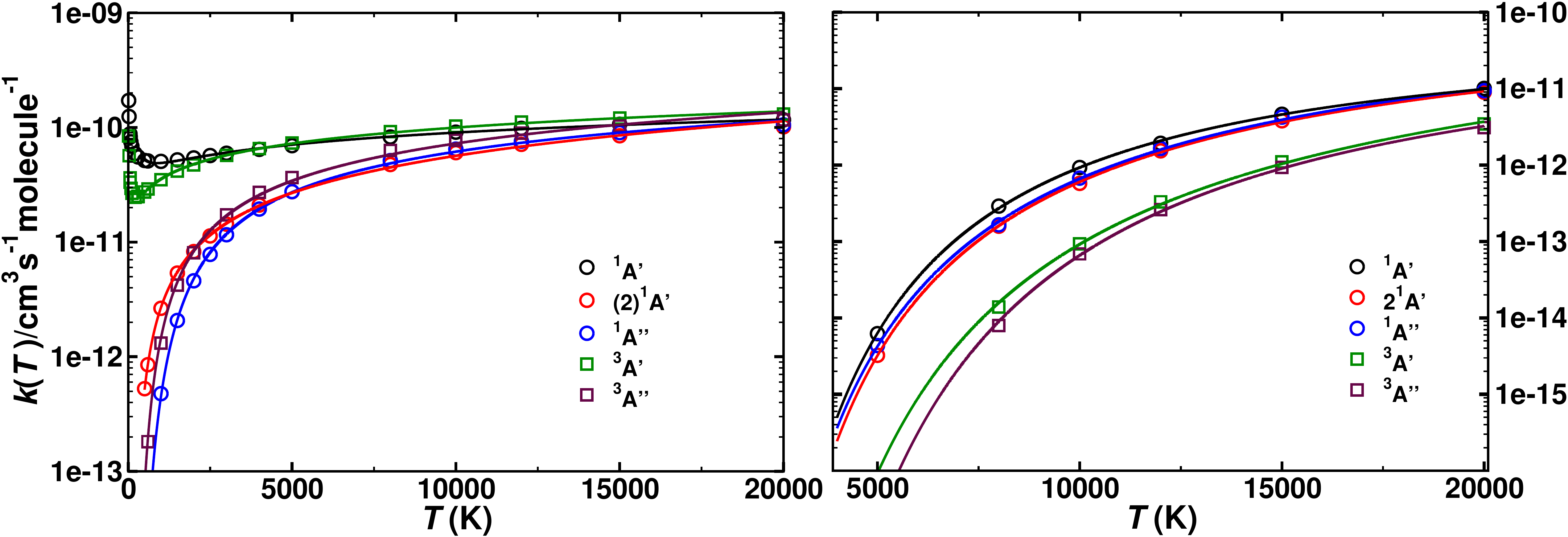}
\caption{Rates for the forward (left) and reverse (right) reactions
  involving all five PESs. Symbols are the QCT results and solid lines
  are the fits to the modified Arrhenius equation. Circles for the
  rates on the singlet PESs and squares for those on the triplet
  PESs. Note the different scales along the $y-$axis for the forward
  and reverse rates.}
\label{sifig:allrates}
\end{figure}

\begin{figure}[h]
    \centering
    \includegraphics[scale=0.6]{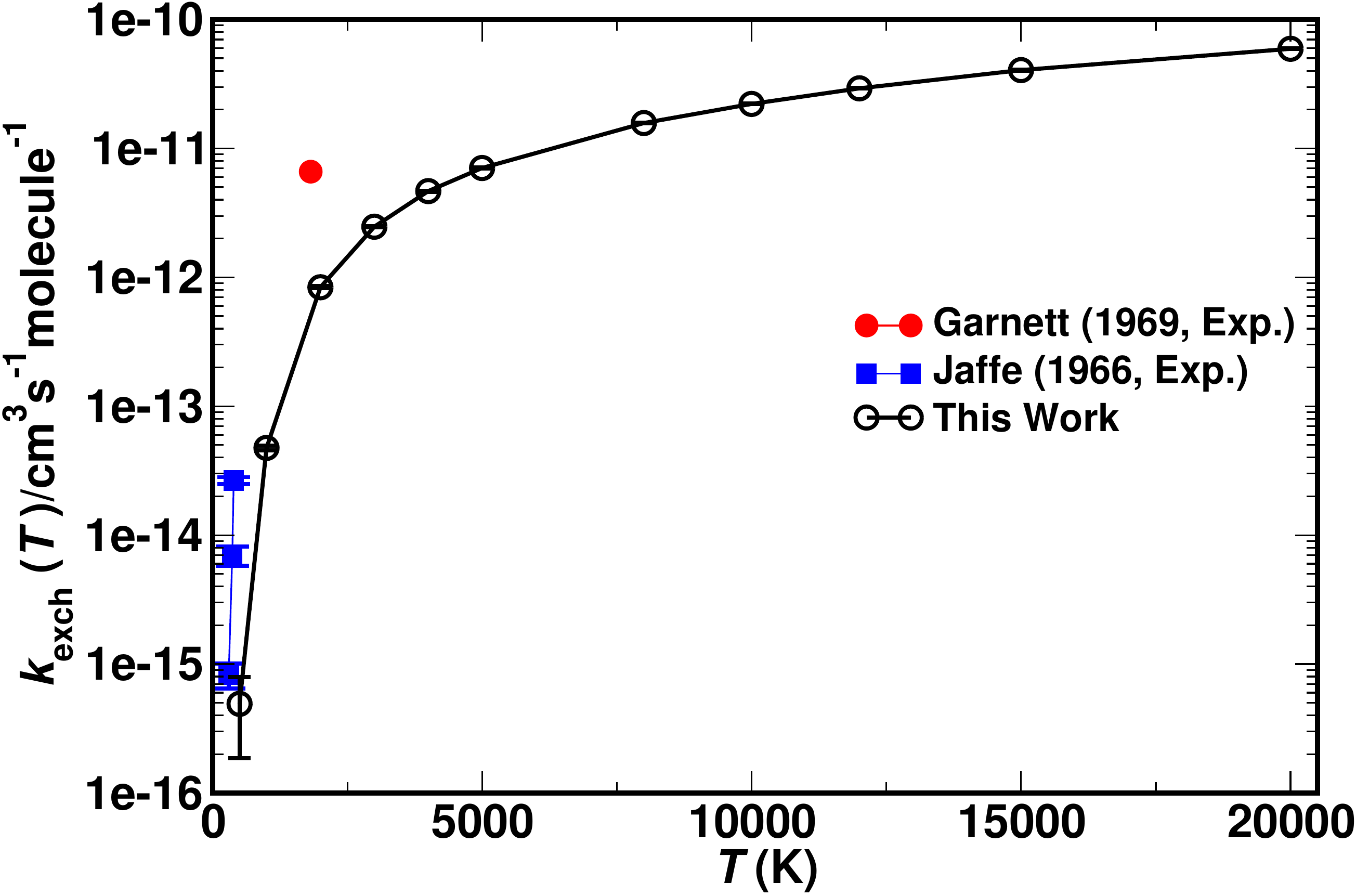}
    \caption{Comparison of computed and experimentally
      observed\cite{Jaffe1966,Garnett1969} rates for the CO$_{\rm
        A}$($^{1}\Sigma^{+}$)+ O$_{\rm B}$($^{3}$P) $\rightarrow$
      CO$_{\rm B}$($^{1}\Sigma^{+}$)+ O$_{\rm A}$($^{3}$P) exchange
      reaction. At low temperatures\cite{Jaffe1966} the computations are in better
      agreement with experiment than for the only high-temperature
      measurement.\cite{Garnett1969}}
\label{sifig:kexch}
\end{figure}

\begin{figure}[h!]
    \centering \includegraphics[scale=0.90]{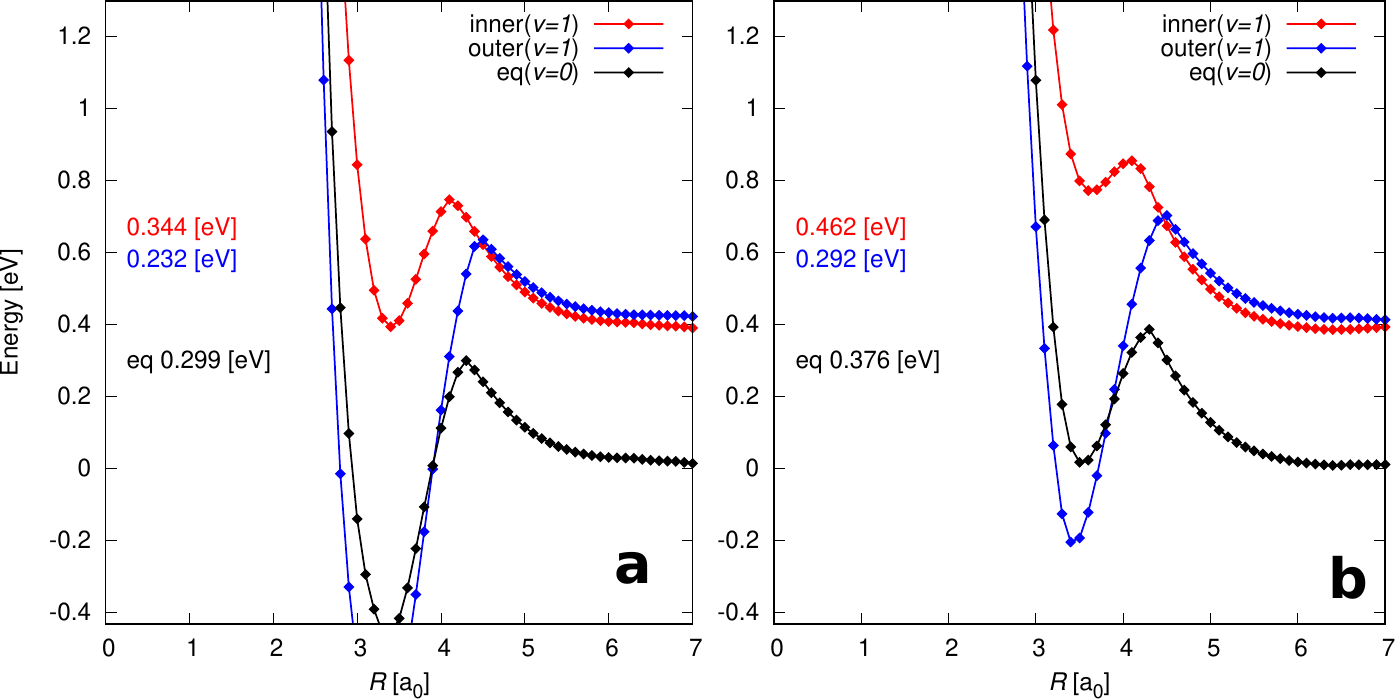}
    \caption{One-dimensional potential energy curves for the CO$_{\rm
        A}$($^{1}\Sigma^{+}$) + O$_{\rm B}$($^{3}$P) $\rightarrow$
      CO$_{\rm B}$($^{1}\Sigma^{+}$) + O$_{\rm A}$($^{3}$P) atom
      exchange reaction on the $^3$A$'$ (left) and $^3$A$''$ PESs. The
      curves are drawn for fixed CO bond length at the equilibrium
      (black), ($v=1$) inner (red), and ($v=1$) outer (blue) turning
      points. For both electronic states the barrier height
      experienced by the approaching oxygen atom O$_{\rm B}$ depends
      on the separation of the CO$_{\rm A}$ diatom.}
\label{sifig:barrierscan}
\end{figure}

\begin{figure}[h!]
    \centering
    \includegraphics[scale=0.70]{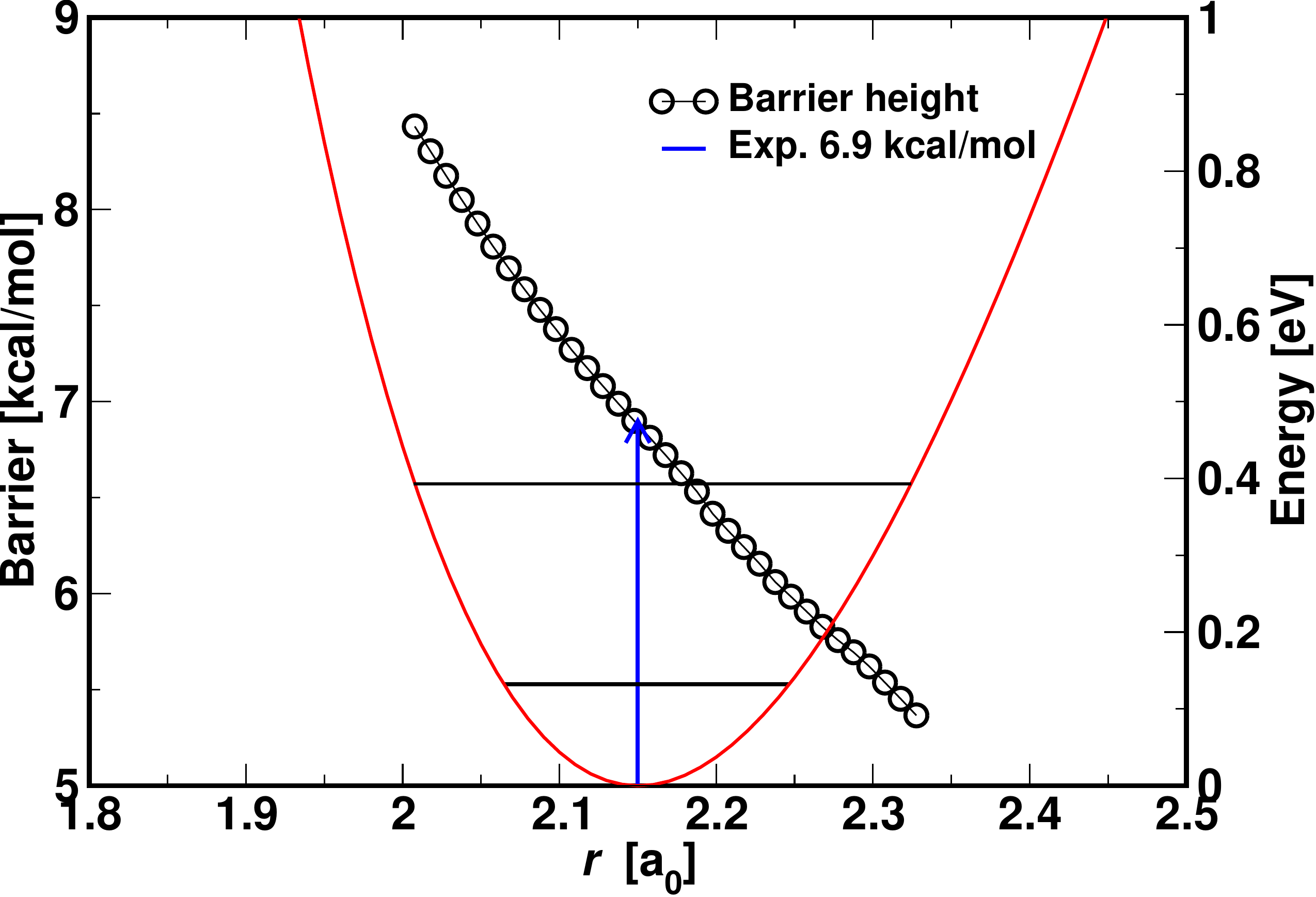}
    \caption{The barrier for the CO$_{\rm A}$($^{1}\Sigma^{+}$) +
      O$_{\rm B}$($^{3}$P) $\rightarrow$ CO$_{\rm
        B}$($^{1}\Sigma^{+}$) + O$_{\rm A}$($^{3}$P) atom exchange
      reaction on the $^3$A$'$ PES evaluated for CO separations
      between the inner and outer turning points for $v_{\rm CO} =
      1$. The CO potential with the $v=0$ and $v=1$ state is
      superimposed. The figure clarifies that the barrier towards
      formation of CO$_2$ varies between 8.5 kcal/mol (0.369 eV) and
      5.5 kcal/mol (0.239 eV) at the inner and outer $v=1$ turning
      points. Hence, the CO vibration acts as a gating mode. The
      barrier height for the isotopic exchange reaction at 300 K was
      reported\cite{Jaffe1966} to be 6.9 kcal/mol (0.299 eV),
      consistent with the barriers found here.}
\label{sifig:gating}
\end{figure}

\begin{figure}
\includegraphics[scale=0.8]{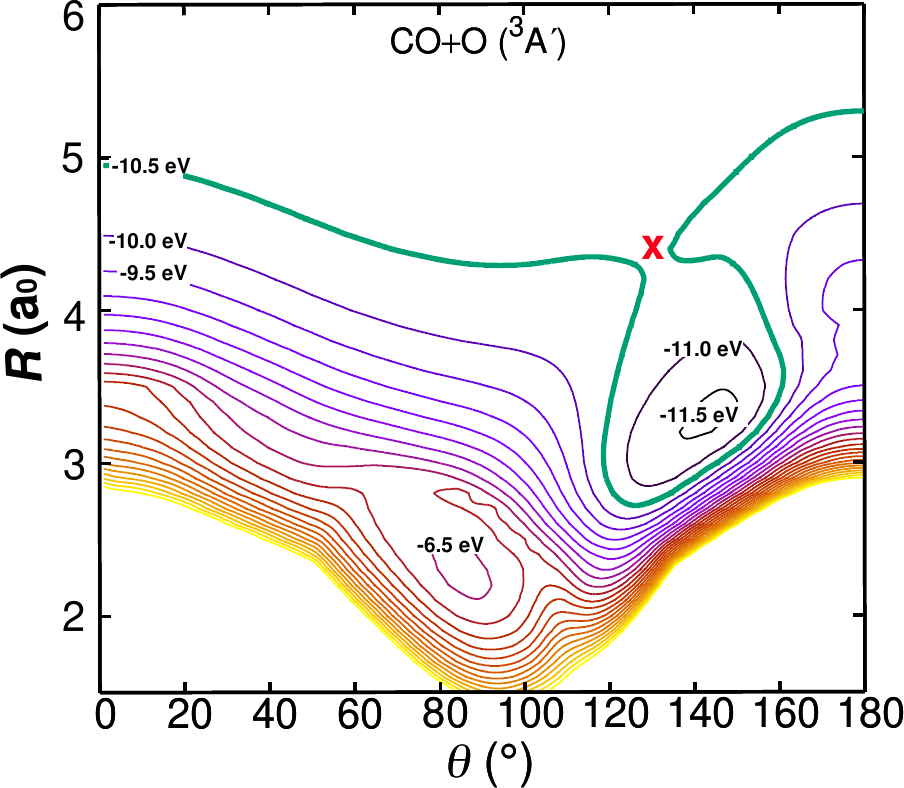}
\caption{Relaxed 2D-PES for the CO + O asymptote for the $^3$A$'$
  state. The diatomic distance $r$ ranges from 2.1 a$_0$ to 3.0
  a$_0$. Contour lines are separated by 0.5 eV between 0 and --11.5 eV
  with the C+O+O dissociation energy as the zero of energy. The solid
  green line represents the --10.5 eV isocontour as a reference. It is
  noted that the relaxing and reactive trajectories leave the strongly
  interacting region through the opening, indicated by the red cross,
  around $R \sim 4.0$ a$_0$ and $\theta = 130^\circ$.}
\label{sifig:pes.relax}
\end{figure}

\begin{figure}[h]
    \centering
    \includegraphics[scale=0.6]{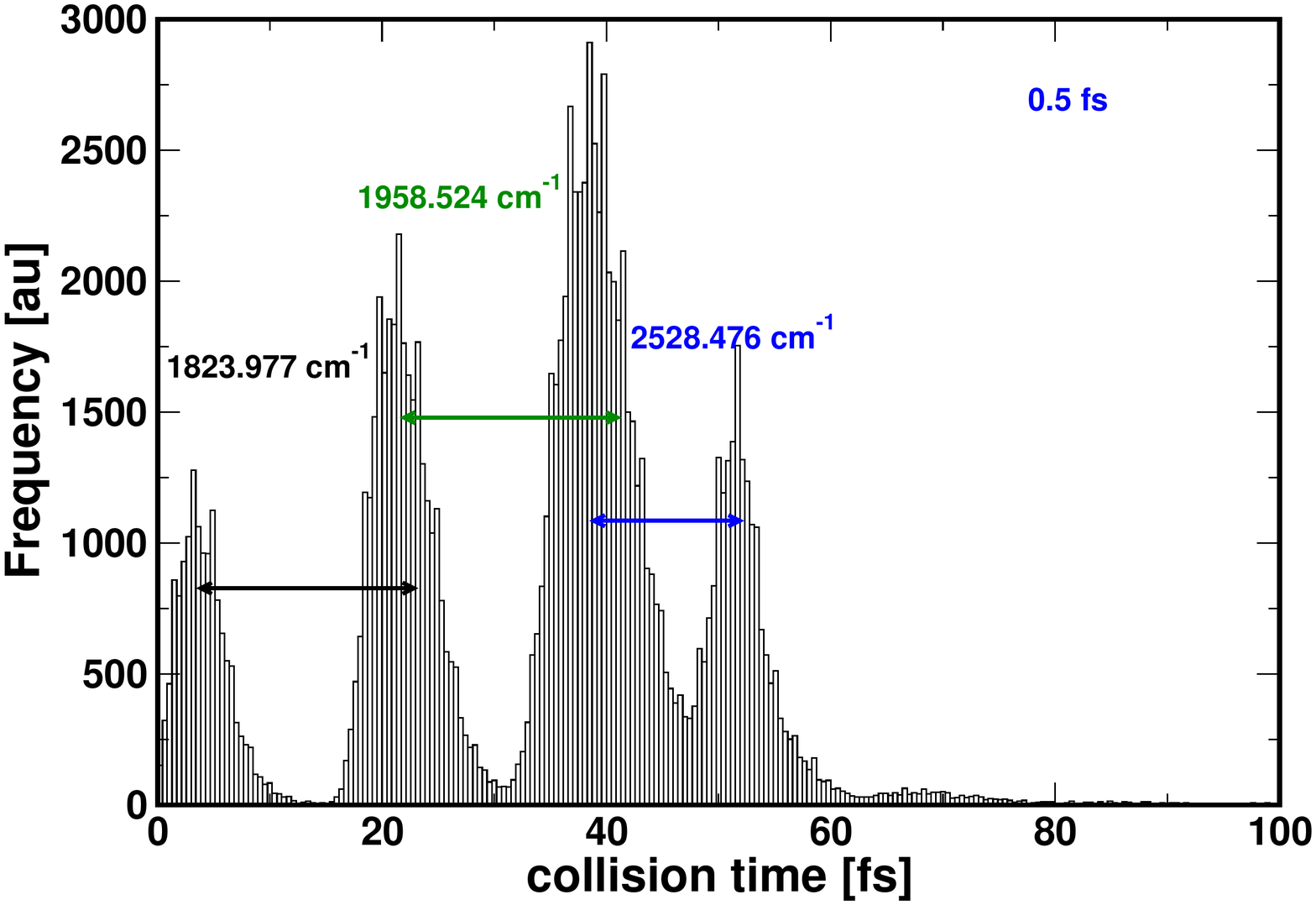}
    \caption{Collision time histogram for $\nu=1$. Difference between
      adjacent peaks are displayed in cm$^{-1}$}
\label{sifig:vr.individ}
\end{figure}

\begin{figure}[h!]
    \centering
    \includegraphics[scale=0.8]{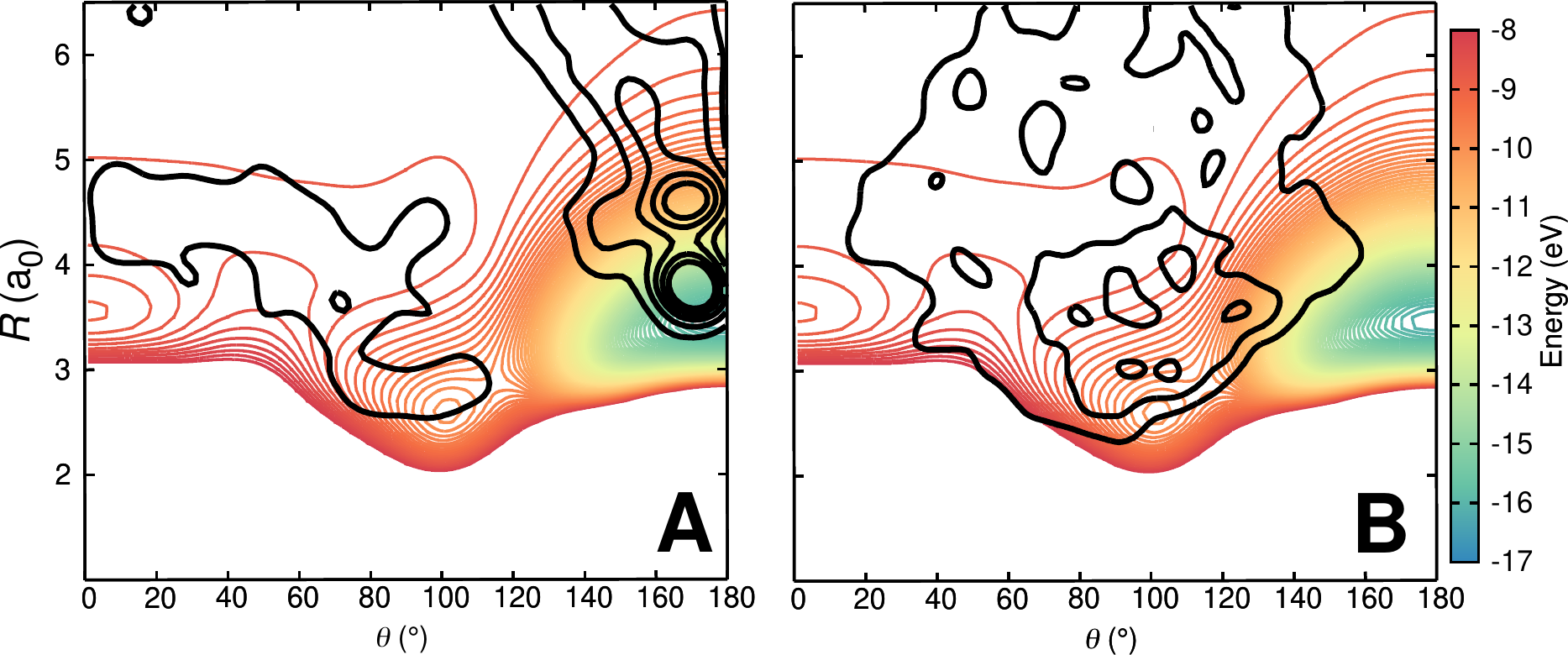}
    \caption{Density trajectory map at 15 K (panel A) and 10000 K
      (panel B) for the C+O$_{2} \longrightarrow$ CO+O($^1$D) reactive
      collisions on the ground state $^{1}$A$'$ PES. The density map
      for the trajectories is superimposed on a relaxed 2D RKHS PES
      where $2.00<r<2.30$ a$_{0}$ (turning points). 300 reacting
      trajectories were taken for each case and represented as a
      KDE. It is found that although both sets of trajectories
      describe the same physical process (atom exchange reaction) they
      are sensitive to and sample different parts of the PES. }
\label{sifig:sample.temp}
\end{figure}

\begin{figure}[h!]
    \centering
    \includegraphics[scale=0.30]{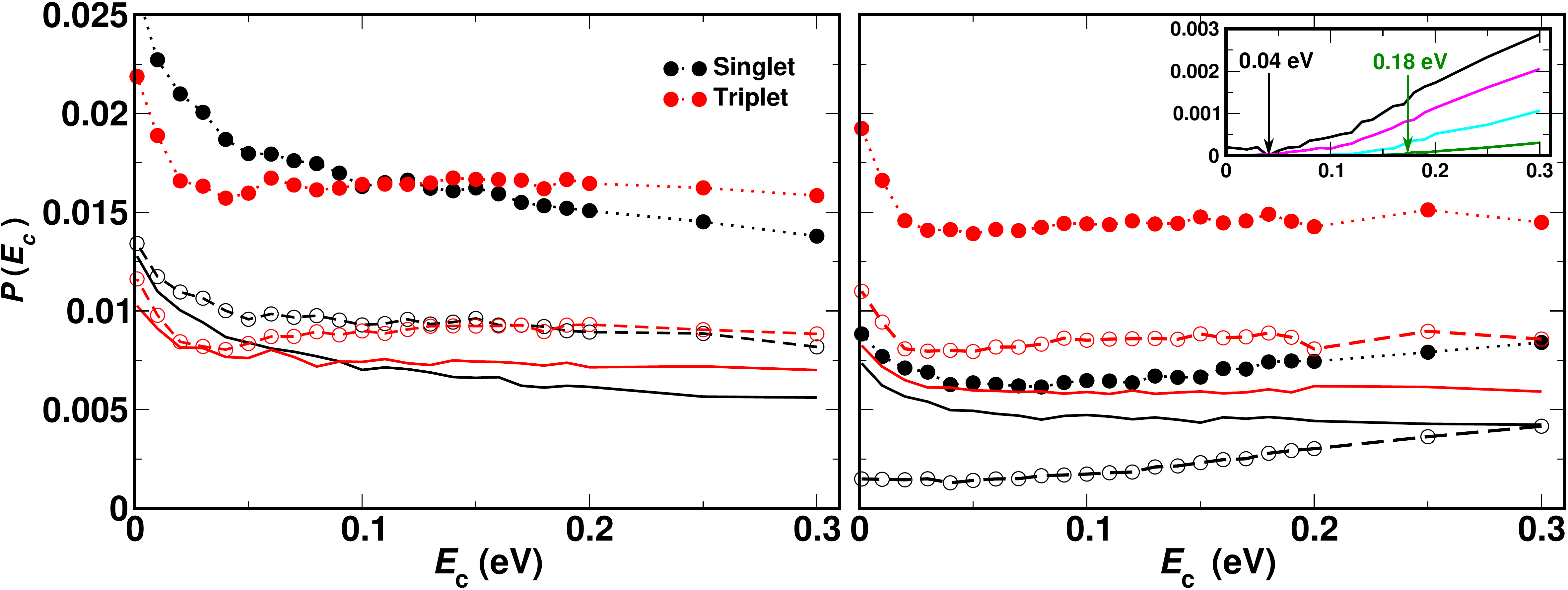}
 \caption{Reaction probability $P(E_{\rm c})$ as a function of
   collision energy $E_{\rm c}$ for the C($^3$P) + O$_{2}$($^3
   \Sigma_{g}^{-})$ $\longrightarrow$ CO($^{1}\Sigma^{+}$) +
   O($^{1}$D) (singlet (black), $^{1}$A$'$ PES) and C($^3$P) +
   O$_{2}$($^3 \Sigma_{g}^{-})$ $\longrightarrow$
   CO($^{1}\Sigma^{+}$) + O($^{3}$P) (triplet (red), $^{3}$A$'$ PES)
   reaction. Left panel for final vibrational state CO ($v' = 16$) and
   right panel for CO($v' = 17$). At least $10^5$ trajectories were
   run for each collision energy (0.001-0.300 eV). Filled circles:
   total reaction probability; open circles: reaction probability
   originating from O$_2$($v = 0$); solid line: reaction probability
   originating from O$_2$($v > 0$). The inset in the right hand panel
   shows an enlargement for the singlet channel (forming O($^1$D)) and
   highlights the threshold energy observed from
   experiment\cite{costes:1998} to open this product channel at
   $E_{\rm c} = 0.04$ eV. The green, cyan, magenta, and black traces
   correspond to initial O$_2 (v=0, j_{\rm max} < 10)$, O$_2 (v=0,
   j_{\rm max} < 20)$, O$_2 (v=0, j_{\rm max} < 30)$, and O$_2 (v=0)$
   for all $j-$values.}
\label{sifig:vibfinal}
\end{figure}

\begin{figure}
  \begin{center}
\includegraphics[scale=0.6]{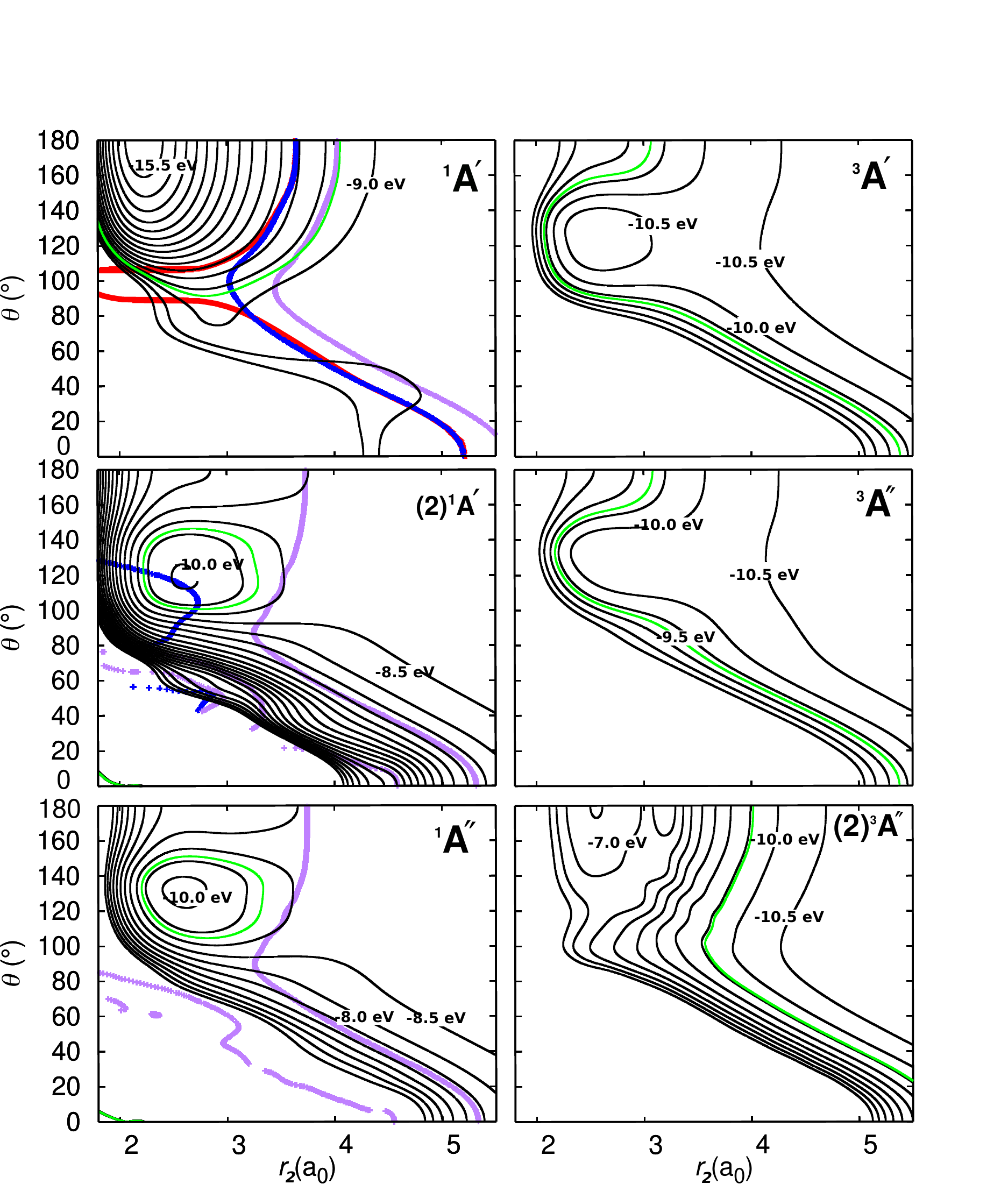}
\caption{Crossings between the singlet and triplet PESs at the inner
  turning point ($r_{1} = 2.06$ a$_0$) of the CO diatom. The distance
  $r_{2}$ is the second CO distance. Contour representation of the
  singlet (left) and triplet (right) states. Colour lines represent
  the intersection seam between the singlet and respective triplet
  surfaces.  $^1$X--$^3$A$'$ (red), $^1$X--$^3$A$''$ (blue) and
  $^1$X--$(2)^3$A$''$ (magenta). Green line indicates a reference
  contour at --9.5 eV relative to the O+O+C dissociation.}
\label{sifig:figcross1}
\end{center}
\end{figure}

\begin{figure}
  \begin{center}
\includegraphics[scale=0.6]{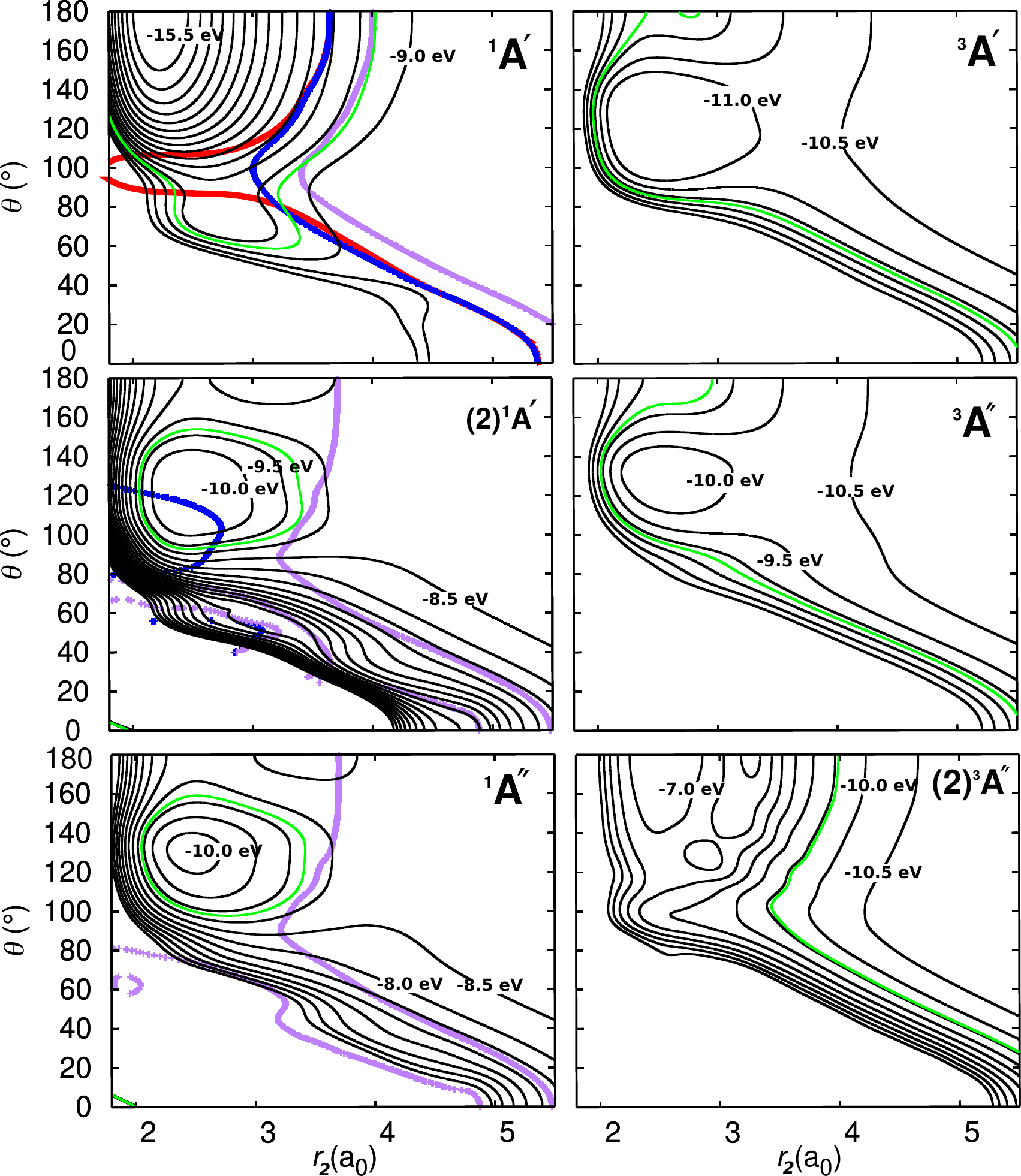}
\caption{Crossings between the PESs at the outer turning point ($r =
  2.24$ a$_0$) of the CO diatom. The distance $r_{2}$ is the second CO
  distance.  Contour representation of the singlet (left) and triplet
  (right) states. Colour lines represent the intersection seam between
  the singlet and respective triplet surfaces.  $^1$X--$^3$A$'$ (red),
  $^1$X--$^3$A$''$ (blue) and $^1$X--$(2)^3$A$''$ (magenta). Green
  line indicates a reference contour at --9.5 eV relative to the O+O+C
  dissociation.}
\label{sifig:figcross2}
\end{center}
\end{figure}

\begin{table}[h!]
\caption{Rates for the C($^3$P) + O$_2$($^3\Sigma_{g}^{-}$)
  $\rightarrow$ CO($^{1}\Sigma^{+}$) + O($^1$D$_{2}$) from 15 to
  20000 K calculated using QCT on the $^1$A$'$, $(2)^1$A$'$, $^1$A$''$
  PESs. Units for rates are in cm$^3$s$^{-1}$molecule$^{-1}$.  $N_{r}$
  is the number of reacting trajectories. }
	\begin{center}
	\begin{tabular}{cccccc}
	\hline
	\hline
\ \ $T$ (K) \ \ & \ \ $N_{r}$ \ \  & \ \  $k^f_1(^1A')$ \ \ \  & \ \ \ $k^f_2((2)^1A')$ \ \ \  & \ \ \  $k^f_3(^1A'')$ \ \ \ & \ \ \  $k^f_1(^1A')$ + $k^f_2((2)^1A')$ + $k^f_3(^1A'')$ \\
	\hline
15 & 93057      & $1.718 \times 10^{-10}$  &  0.000  &  0.000  &  $1.718 \times 10^{-10}$  \\
27     & 90583  & $1.245 \times 10^{-10}$  &  0.000  &  0.000  &  $1.245 \times 10^{-10}$  \\
54     & 85791  & $8.778 \times 10^{-11}$  &  0.000  &  0.000  &  $8.778 \times 10^{-11}$  \\
63     & 84463  & $8.261 \times 10^{-11}$  &  0.000  &  0.000  &  $8.261 \times 10^{-11}$  \\
83     & 81512  & $7.501 \times 10^{-11}$  &  0.000  &  0.000  &  $7.501 \times 10^{-11}$  \\
112    & 77896  & $6.868 \times 10^{-11}$  &  $2.617 \times 10^{-17}$  &  0.000  &  $6.868 \times 10^{-11}$  \\
207    & 68861  & $5.925 \times 10^{-11}$  &  $1.186 \times 10^{-14}$  &  0.000  &  $5.926 \times 10^{-11}$  \\
295    & 66814  & $5.551 \times 10^{-11}$  &  $7.371 \times 10^{-14}$  &  $2.095 \times 10^{-16}$  &  $5.558 \times 10^{-11}$  \\
300    & 66493  & $5.532 \times 10^{-11}$  &  $7.785 \times 10^{-14}$  &  $2.108 \times 10^{-16}$  &  $5.540 \times 10^{-11}$  \\
500    & 58788  & $5.152 \times 10^{-11}$  &  $5.265 \times 10^{-13}$  &  $6.183 \times 10^{-15}$  &  $5.205 \times 10^{-11}$  \\
600    & 56358  & $5.104 \times 10^{-11}$  &  $8.512 \times 10^{-13}$  &  $2.394 \times 10^{-14}$  &  $5.191 \times 10^{-11}$  \\
1000   & 53536  & $5.057 \times 10^{-11}$  &  $2.643 \times 10^{-12}$  &  $4.761 \times 10^{-13}$  &  $5.369 \times 10^{-11}$  \\
1500   & 49748  & $5.217 \times 10^{-11}$  &  $5.368 \times 10^{-12}$  &  $2.070 \times 10^{-12}$  &  $5.961 \times 10^{-11}$  \\
2000   & 50708  & $5.421 \times 10^{-11}$  &  $8.266 \times 10^{-12}$  &  $4.592 \times 10^{-12}$  &  $6.707 \times 10^{-11}$  \\
2500   & 49341  & $5.686 \times 10^{-11}$  &  $1.132 \times 10^{-11}$  &  $7.810 \times 10^{-12}$  &  $7.599 \times 10^{-11}$  \\
3000   & 48425  & $5.955 \times 10^{-11}$  &  $1.434 \times 10^{-11}$  &  $1.159 \times 10^{-11}$  &  $8.549 \times 10^{-11}$  \\
4000   & 46964  & $6.442 \times 10^{-11}$  &  $2.102 \times 10^{-11}$  &  $1.942 \times 10^{-11}$  &  $1.049 \times 10^{-10}$  \\
5000   & 46270  & $6.939 \times 10^{-11}$  &  $2.752 \times 10^{-11}$  &  $2.757 \times 10^{-11}$  &  $1.245 \times 10^{-10}$  \\
8000   & 52737  & $8.250 \times 10^{-11}$  &  $4.748 \times 10^{-11}$  &  $5.138 \times 10^{-11}$  &  $1.814 \times 10^{-10}$  \\
10000  & 52363  & $9.090 \times 10^{-11}$  &  $6.025 \times 10^{-11}$  &  $6.469 \times 10^{-11}$  &  $2.158 \times 10^{-10}$  \\
12000  & 51930  & $9.818 \times 10^{-11}$  &  $7.120 \times 10^{-11}$  &  $7.585 \times 10^{-11}$  &  $2.452 \times 10^{-10}$  \\
15000  & 50987  & $1.066 \times 10^{-10}$  &  $8.462 \times 10^{-11}$  &  $9.010 \times 10^{-11}$  &  $2.813 \times 10^{-10}$  \\
20000  & 49054  & $1.164 \times 10^{-10}$  &  $1.013 \times 10^{-10}$  &  $1.054 \times 10^{-10}$  &  $3.232 \times 10^{-10}$  \\
	\hline
	\hline
  \end{tabular}
	\end{center}
\label{sitab:fwd.singlet}
\end{table}

\begin{table}[h!]
\caption{Rate coefficients for the C($^3$P) +
  O$_2$($^3\Sigma_{g}^{-}$) $\rightarrow$ CO($^{1}\Sigma^{+}$) +
  O($^3$P) from 15 to 20000 K calculated using QCT on the $^3$A$'$ and
  $^3$A$''$ PESs. Units are in cm$^3$s$^{-1}$molecule$^{-1}$.  $N_{r}$
  is the number of reacting trajectories. }
	\begin{center}
	\begin{tabular}{ccccc}
	\hline
	\hline
\ \ $T$ (K) \ \ \ & \ \ \  $N_{r}$ \ \ \ &  \ \ \ $k^f_1(^3A')$ \ \ \ & \ \ \ $k^f_2(^3A'')$ \ \ \ \ &  \ \ \  $k^f_1(^3A')$  + $k^f_2(^3A'')$ \\
	\hline
15     &  49736  &  $8.414 \times 10^{-11}$  &  0.000  &  $8.414 \times 10^{-11}$  \\
27     &  47390  &  $5.682 \times 10^{-11}$  &  0.000  &  $5.682 \times 10^{-11}$  \\
54     &  44900  &  $3.632 \times 10^{-11}$  &  0.000  &  $3.632 \times 10^{-11}$  \\
63     &  44406  &  $3.337 \times 10^{-11}$  &  0.000  &  $3.337 \times 10^{-11}$  \\
83     &  43644  &  $2.949 \times 10^{-11}$  &  0.000  &  $2.949 \times 10^{-11}$  \\
112    &  43489  &  $2.671 \times 10^{-11}$  &  0.000  &  $2.671 \times 10^{-11}$  \\
207    &  43238  &  $2.460 \times 10^{-11}$  &  0.000  &  $2.460 \times 10^{-11}$  \\
295    &  46131  &  $2.486 \times 10^{-11}$  &  $1.467 \times 10^{-15}$  &  $2.487 \times 10^{-11}$  \\
300    &  46187  &  $2.506 \times 10^{-11}$  &  $1.476 \times 10^{-15}$  &  $2.506 \times 10^{-11}$  \\
500    &  46573  &  $2.742 \times 10^{-11}$  &  $8.043 \times 10^{-14}$  &  $2.750 \times 10^{-11}$  \\
600    &  46856  &  $2.906 \times 10^{-11}$  &  $1.818 \times 10^{-13}$  &  $2.924 \times 10^{-11}$  \\
1000   &  50249  &  $3.498 \times 10^{-11}$  &  $1.321 \times 10^{-12}$  &  $3.630 \times 10^{-11}$  \\
1500   &  50968  &  $4.156 \times 10^{-11}$  &  $4.213 \times 10^{-12}$  &  $4.578 \times 10^{-11}$  \\
2000   &  54396  &  $4.706 \times 10^{-11}$  &  $8.086 \times 10^{-12}$  &  $5.514 \times 10^{-11}$  \\
3000   &  54474  &  $5.717 \times 10^{-11}$  &  $1.729 \times 10^{-11}$  &  $7.446 \times 10^{-11}$  \\
4000   &  54387  &  $6.554 \times 10^{-11}$  &  $2.710 \times 10^{-11}$  &  $9.264 \times 10^{-11}$  \\
5000   &  54311  &  $7.304 \times 10^{-11}$  &  $3.655 \times 10^{-11}$  &  $1.096 \times 10^{-10}$  \\
8000   &  62916  &  $9.208 \times 10^{-11}$  &  $6.297 \times 10^{-11}$  &  $1.551 \times 10^{-10}$  \\
10000  &  62254  &  $1.025 \times 10^{-10}$  &  $7.745 \times 10^{-11}$  &  $1.799 \times 10^{-10}$  \\
12000  &  61260  &  $1.113 \times 10^{-10}$  &  $8.944 \times 10^{-11}$  &  $2.007 \times 10^{-10}$  \\
15000  &  59333  &  $1.205 \times 10^{-10}$  &  $1.027 \times 10^{-10}$  &  $2.232 \times 10^{-10}$  \\
20000  &  55854  &  $1.306 \times 10^{-10}$  &  $1.166 \times 10^{-10}$  &  $2.472 \times 10^{-10}$  \\
	\hline
	\hline
  \end{tabular}
	\end{center}
\label{sitab:fwd.triplet}
\end{table}

\begin{table}[h!]
  \caption{Rate coefficients for the CO($^{1}\Sigma^{+}$)+ O($^{1}$D)
    $\rightarrow$ C($^3$P) + O$_2$($^3\Sigma_{g}^{-}$) from 3000
    to 20000 K calculated using QCT on the $^1$A$'$, $(2)^1$A$'$,
    $^1$A$''$ PESs. Units are in cm$^3$s$^{-1}$molecule$^{-1}$.
    $N_{r}$ is the number of reacting trajectories.}
	\begin{center}
	\begin{tabular}{cccccc}
	\hline
	\hline
\ \ $T$ (K) \ \ \ & \ \ \  $N_{r}$ \ \ \ &  \ \ \ $k^r_1(^1A')$ \ \ \ & \ \ \ $k^r_2((2)^1A')$ \ \ \ \ & \ \ \ \ $k^r_3(^1A'')$ \ \ \ & \ \ \  $k^r_1(^1A')$ + $k^r_2((2)^1A')$ + $k^r_3(^1A'')$ \\
	\hline
3000   & 0     & 0.000  &  0.000  &  0.000  &  0.000  \\
4000   & 3     & $6.226 \times 10^{-16}$  &  $3.738 \times 10^{-16}$  &  0.000  &  $9.964 \times 10^{-16}$  \\
5000   & 16    & $6.236 \times 10^{-15}$  &  $3.239 \times 10^{-15}$  &  $4.351 \times 10^{-15}$  &  $1.383 \times 10^{-14}$  \\
8000   & 383   & $2.902 \times 10^{-13}$  &  $1.588 \times 10^{-13}$  &  $1.651 \times 10^{-13}$  &  $6.141 \times 10^{-13}$  \\
10000  & 1037  & $9.278 \times 10^{-13}$  &  $5.759 \times 10^{-13}$  &  $6.744 \times 10^{-13}$  &  $2.178 \times 10^{-12}$  \\
12000  & 1897  & $1.928 \times 10^{-12}$  &  $1.533 \times 10^{-12}$  &  $1.653 \times 10^{-12}$  &  $5.114 \times 10^{-12}$  \\
15000  & 3711  & $4.616 \times 10^{-12}$  &  $3.796 \times 10^{-12}$  &  $4.239 \times 10^{-12}$  &  $1.265 \times 10^{-11}$  \\
20000  & 6655  & $1.008 \times 10^{-11}$  &  $8.983 \times 10^{-12}$  &  $9.373 \times 10^{-12}$  &  $2.844 \times 10^{-11}$  \\
	\hline
	\hline
  \end{tabular}
	\end{center}
\label{sitab:rev.singlet}
\end{table}

\begin{table}[h!]
  \caption{Rate coefficients for the CO($^{1}\Sigma^{+}$)+ O($^{3}$P)
    $\rightarrow$ C($^3$P) + O$_2$($^3\Sigma_{g}^{-}$) from 3000
    to 20000 K calculated using QCT on the $^3$A$'$ and
    $^3$A$''$ PESs. Units are in cm$^3$s$^{-1}$molecule$^{-1}$.
    $N_{r}$ is the number of reacting trajectories.}
	\begin{center}
	\begin{tabular}{ccccc}
	\hline
	\hline
\ \ $T$ (K) \ \ \ & \ \ \  $N_{r}$ \ \ \ &  \ \ \ $k^r_1(^3A')$ \ \ \ & \ \ \ $k^r_2(^3A'')$ \ \ \ \ &  \ \ \  $k^r_1(^3A')$  + $k^r_2(^3A'')$ \\
	\hline
3000  &  0     &  0.000  &  0.000  &  0.000  \\	
4000  &  0     &  0.000  &  0.000  &  0.000  \\
5000  &  2     &  $9.055 \times 10^{-17}$  &  $2.175 \times 10^{-17}$  &  $1.123 \times 10^{-16}$  \\
8000  &  223   &  $1.429 \times 10^{-14}$  &  $7.940 \times 10^{-15}$  &  $2.223 \times 10^{-14}$  \\
10000 &  1332  &  $9.277 \times 10^{-14}$  &  $6.892 \times 10^{-14}$  &  $1.617 \times 10^{-13}$  \\
12000 &  4339  &  $3.340 \times 10^{-13}$  &  $2.609 \times 10^{-13}$  &  $5.948 \times 10^{-13}$  \\
15000 &  11796 &  $1.098 \times 10^{-12}$  &  $9.348 \times 10^{-13}$  &  $2.033 \times 10^{-12}$  \\
20000 &  28176 &  $3.449 \times 10^{-12}$  &  $3.087 \times 10^{-12}$  &  $6.535 \times 10^{-12}$  \\
	\hline
	\hline
  \end{tabular}
	\end{center}
\label{sitab:rev.triplet}
\end{table}

\begin{table}[h!]
  \caption{Temperature dependent rates for the CO$_{\rm
      A}$($^{1}\Sigma^{+}$)+ O$_{\rm B}$($^{3}$P) $\rightarrow$
    CO$_{\rm B}$($^{1}\Sigma^{+}$)+ O$_{\rm A}$($^{3}$P) exchange
    reaction from 500 to 20000 K calculated using QCT on the $^3$A$'$
    PESs. Units are in cm$^3$s$^{-1}$molecule$^{-1}$.  $N_{r}$ is the
    number of reacting trajectories.}
\begin{center}
\begin{tabular}{cccccc}
\hline
\hline
\ \ $T$ (K) \ \ \ & \ \ \   $N$ \ \ \ & \ \ \ \ $k^f_1(^3A')$ \ \ \ \  & \ \ \ \ $k^f_1(^3A'')$ \ \ \ \  & \ \ \ \ $k^f_1(^3A'+^3A'')$ \ \ \ \ & \ \ \ $\Delta k^f_1(^3A'+^3A'')$ \\
\hline
500    & 57     & $3.859 \times 10^{-16}$   &   $1.053  \times 10^{-16}$   &   $4.912   \times 10^{-16}$  & $3.052 \times 10^{-16}$ \\
1000   & 425    & $3.511 \times 10^{-14}$   &   $1.242  \times 10^{-14}$   &   $4.754   \times 10^{-14}$  & $1.972 \times 10^{-15}$ \\
2000   & 4734   & $5.570 \times 10^{-13}$   &   $2.812  \times 10^{-13}$   &   $8.382   \times 10^{-13}$  & $1.746 \times 10^{-14}$ \\
3000   & 11621  & $1.503 \times 10^{-12}$   &   $9.694  \times 10^{-13}$   &   $2.472   \times 10^{-12}$  & $3.340 \times 10^{-14}$ \\
4000   & 20393  & $2.649 \times 10^{-12}$   &   $2.012  \times 10^{-12}$   &   $4.661   \times 10^{-12}$  & $3.969 \times 10^{-14}$ \\
5000   & 28143  & $3.808 \times 10^{-12}$   &   $3.233  \times 10^{-12}$   &   $7.041   \times 10^{-12}$  & $5.890 \times 10^{-14}$ \\
8000   & 52280  & $7.947 \times 10^{-12}$   &   $7.777  \times 10^{-12}$   &   $1.572   \times 10^{-11}$  & $7.493 \times 10^{-14}$ \\
10000  & 78807  & $1.116 \times 10^{-11}$   &   $1.099  \times 10^{-11}$   &   $2.215   \times 10^{-11}$  & $1.088 \times 10^{-13}$ \\
12000  & 93891  & $1.459 \times 10^{-11}$   &   $1.468  \times 10^{-11}$   &   $2.927   \times 10^{-11}$  & $1.606 \times 10^{-13}$ \\
15000  & 127016 & $2.008 \times 10^{-11}$   &   $2.043  \times 10^{-11}$   &   $4.051   \times 10^{-11}$  & $3.037 \times 10^{-13}$ \\
20000  & 182669 & $2.932 \times 10^{-11}$   &   $3.001  \times 10^{-11}$   &   $5.933   \times 10^{-11}$  & $1.990 \times 10^{-13}$ \\ 
\hline
\hline
\end{tabular}
\end{center}
\label{sitab:kexch}
\end{table}

\begin{table}[h!]
  \caption{Rate coefficients for the CO($^{1}\Sigma^{+}$)+ O($^{3}$P)
    $\rightarrow$ CO($^{1}\Sigma^{+}$)+ O($^{3}$P) vibrational
    relaxation $\nu=1 \rightarrow 0$ from 300 to 5000 K calculated
    using QCT on the $^3$A$'$ and $^3$A$''$ PESs. Units for rates are
    in cm$^3$s$^{-1}$molecule$^{-1}$ and $N_{r}$ is the number of
    reacting trajectories.}
	\begin{center}
	\begin{tabular}{cccccc}
	\hline
	\hline
\ \ $T$ (K) \ \ \ & \ \ \ \  $N_{r}$ \ \ \ &  \ \ $k^f_1(^3A')$ \ \ \ \ \ \ & \ \ \ \ \ \ $k^f_1(^3A'')$ \ \ \ & \ \ \  $k^f_1(^3A')$ + $k^f_1(^3A'')$ \\
	\hline
300    & 0    & 0.000  &  0.000  &  0.000  \\
500    & 5    & $1.086 \times 10^{-15}$  &  $9.675 \times10^{-16}$  &  $3.021 \times 10^{-15}$  \\
1000   & 301  & $1.235 \times 10^{-13}$  &  $1.081 \times10^{-13}$  &  $3.396 \times 10^{-13}$  \\
2000   & 1713 & $1.329 \times 10^{-12}$  &  $1.490 \times10^{-12}$  &  $4.309 \times 10^{-12}$  \\
3000   & 2956 & $3.043 \times 10^{-12}$  &  $3.966 \times10^{-12}$  &  $1.097 \times 10^{-11}$  \\
4000   & 3950 & $4.780 \times 10^{-12}$  &  $6.701 \times10^{-12}$  &  $1.818 \times 10^{-11}$  \\
5000   & 4524 & $6.237 \times 10^{-12}$  &  $9.933 \times10^{-12}$  &  $2.610 \times 10^{-11}$  \\
	\hline
	\hline
  \end{tabular}
\end{center}
\label{sitab:corates4}
\end{table}

\newpage

\bibliography{ref2}